\documentclass[review]{elsarticle}

\usepackage{algorithm}
\usepackage{algorithmicx}
\usepackage[noend]{algpseudocode}
\renewcommand{\algorithmicrequire}{ \textbf{Input:}} %Use Input in the format of Algorithm
\renewcommand{\algorithmicensure}{ \textbf{Output:}} %UseOutput in the format of Algorithm
\usepackage{amssymb}
\usepackage[justification=centering]{caption}
\usepackage{subfigure}
\usepackage{bm}
\usepackage{amsmath,amssymb,amsfonts}
\usepackage{booktabs}
\usepackage{longtable}
\usepackage{diagbox}
\usepackage{tabularx}
\usepackage{color, xcolor}
\usepackage{soul}
\soulregister\ref7
\soulregister\cite7

\journal{Journal of \LaTeX\ Templates}

\begin{document}
%\abovedisplayshortskip=0pt
%\belowdisplayshortskip=0pt
%\abovedisplayskip=0pt
%\belowdisplayskip=0pt
\begin{frontmatter}

\title{Stochastic Gradient-based Fast Distributed Multi-Energy  Management for an Industrial Park with Temporally-Coupled Constraints\tnoteref{mytitlenote}}
\tnotetext[mytitlenote]{This work was supported by the National Key R\&D Program of China (Grant No.2016YFB0901900), and in part by the NSF of China (Grants No. 61731012, 61573245, 61922058, and 61973264).}

\author[mymainaddress,mysecondaryaddress]{Dafeng Zhu}

\author[mymainaddress,mysecondaryaddress]{Bo Yang\corref{mycorrespondingauthor}}
\cortext[mycorrespondingauthor]{Corresponding author}
\ead{bo.yang@sjtu.edu.cn}

\author[mythirdaddress]{Chengbin Ma}
\author[mymainaddress,mysecondaryaddress]{Zhaojian Wang}
\author[mymainaddress,mysecondaryaddress]{Shanying Zhu}
\author[myfourthaddress]{Kai Ma}
%\author[myfourthaddress]{Chengbin Ma}
\author[mymainaddress,mysecondaryaddress]{Xinping Guan}

\address[mymainaddress]{Department of Automation, Shanghai Jiao Tong University, Shanghai 200240, China}
\address[mysecondaryaddress]{Key Laboratory of System Control and Information Processing, Ministry of Education of China, Shanghai 200240, China}
\address[mythirdaddress]{University of Michigan-Shanghai Jiao Tong University Joint Institute, Shanghai Jiao Tong University, Shanghai 200240, China}
\address[myfourthaddress]{Key Laboratory of Industrial Computer Control Engineering of Hebei Province, Yanshan University, Qinhuangdao 066004, China}

\begin{abstract}
Contemporary industrial parks are challenged by the growing concerns about high cost and low efficiency of energy supply.  Moreover, in the case of uncertain supply/demand, how to mobilize delay-tolerant elastic loads and compensate real-time inelastic loads to match multi-energy generation/storage and minimize energy cost is a key issue. Since energy management is hardly to be implemented offline without knowing statistical information of random variables, this paper presents a systematic online energy cost minimization framework to fulfill the complementary utilization of multi-energy  with time-varying generation, demand and price. Specifically to achieve charging/discharging constraints due to storage and short-term energy balancing, a fast distributed algorithm based on stochastic gradient with two-timescale implementation is proposed to ensure online implementation. To reduce the peak loads, an incentive mechanism is implemented by estimating users' willingness to shift. Analytical results on parameter setting are also given to guarantee feasibility and optimality of the proposed design. Numerical results show that when the bid-ask spread of electricity is small enough, the proposed algorithm can achieve the close-to-optimal cost asymptotically.
 %of shifting inelastic loads without knowing energy usage of each individual user. To response to the multi-energy charge/discharge, peak load shifting, variable demand, renewable energy and prices without the priori knowledge of the underlying random process, a fast distributed algorithm based on two-stage dual decomposition and stochastic gradient is proposed to deal with temporally-coupled constraints and ensure online implementation.  To further consider the spatiotemporal variation of energy supply and demand, we construct a two-timescale optimization to achieve long-term energy storage balancing and real-time load balancing while respecting energy constraints. Based on the performance analysis and real-data case studies, the proposed algorithm is feasible and asymptotically optimal independent of the statistical characteristics of random events. 
\end{abstract}

\begin{keyword}
Multi-energy industrial park \sep peak loads shifting \sep two-timescale optimization \sep stochastic gradient  \sep fast distributed algorithm 
\end{keyword}

\end{frontmatter}

%\linenumbers

\section{Introduction}
Traditional industrial production burns a large amount of fossil fuels for energy generation, resulting in rapid consumption of fossil fuel and serious pollution. This motivates the research of energy systems about application reliability \cite{Liu2021Bounding}, battery management \cite{Wu2021Battery}, energy balance \cite{Zhang2018Peer}, demand response \cite{X.Zhang} and distributed cooperation \cite{Zhu2022Energy}. Many industrial parks in China have been or are under construction. These parks consume a large amount of electricity provided by power grids. Considering the limited power distribution of each factory, it is necessary to build photovoltaic panels and generators, such as combined heat and power (CHP) units in the parks. %
{In addition, the production of industrial raw materials requires high-temperature and high-pressure steam, which is provided by boilers and CHP units.} With the continuous expansion of production scale and the rapid growth of energy consumption, serious issues such as low energy efficiency and rising operating costs in industrial parks need to be solved urgently. To tackle these problems, energy hubs (EHs) including energy storages, CHP units, boilers and photovoltaic panels, are introduced into an industrial park. By transferring multi-energy supply and demand across time and space, EHs can obtain scheduling flexibility and complementarity, thereby improving energy revenue, reliability and efficiency \cite{Li2020A}. %EHs can obtain the complementarity and flexibility of the schedule by controlling and optimizing the multi-energy networks in different timescales, which can improve the energy reliability and utilization.

A number of studies are conducted on energy management in multi-energy industrial parks to improve energy utilization based on the characteristics of multi-energy. For example, Gu et al. \cite{Wu2021Optimal} establish a bi-level model that considers the gap of peak-valley demands and high penetration of distributed generation to minimize the operation cost of an industrial park, which, however, does not consider renewable energy generation. To quantitatively investigate the relationship between the planning cost and the renewable energy sources, Xu et al. \cite{Xu2020Optimal} establish a demand response model with day-ahead pricing and an allocation method of a multi-energy system in industrial parks. Taking into account the impact of weather factors on the variation in loads and renewable energy, Zhu et al. \cite{Zhu2020Regional} form typical weather scenarios to describe the uncertainty in these factors and propose an energy management strategy of regional integrated energy systems in industrial parks considering correspondence between the multi-energy demand and supply. To accurately evaluate the techno-economic-environmental of industrial parks, Guo et al. \cite{Guo2021Economic} focus on the coordinated operation of parks with electrical, thermal, cooling loads and demand response, which satisfies environmental and economic benefits. 
%In \cite{Chen2020Improving}, a new method using the hydraulic inertia of steam heating network is proposed to improve the flexibility of multi-energy industrial parks. %The demand response is usually used to improve energy efficiency\cite{Jiang2017Interaction,Liu2020Heat,Jiang2019Integrated}.
% In \cite{Jiang2017Interaction}, a universal multi-energy demand response optimization model is established to realize the interaction between industrial users and the power grid based on the output characteristics of CHP. In \cite{Liu2020Heat}, a "synchronous-generation-asynchronous-transmission" management framework for heat and electricity in a multi-energy industrial park is established, and the interaction between CHP unit owners and users are considered to achieve peak load shifting. In \cite{Jiang2019Integrated}, the coupling characteristics of industrial multi-energy demands are modeled to reflect the influence between different energy consumption behaviors. %interaction between supply and demand in multi-energy demand management was discussed, and a demand response method for industrial parks was proposed. 
 %
{However, these studies focus on the interaction of multi-energy supply and demand, and few consider the combination of multi-energy generation, load and the energy storage balance, which is indispensable in the future multi-energy generation-storage-coordinated industrial production.} %demand response and interaction individually, the combination of computational complexity, stochastic nature and convergence rate has not been explored. %for the model of the industrial users, who care the cost but not the source of energy.   

%
%{Although renewable energy generation, energy storage and demand response are effective methods to release the imbalance between supply and demand, the time-varying and stochastic nature of energy generation, storage and demand should be taken into account.} % 
Due to the rapid growth of energy demand for industrial production, some industrial users are facing pressures such as insufficient energy supply, low energy utilization, and overload. To reduce the impact, peak load shifting is considered in energy management. For instance, an energy balance provider is involved to further improve the benefit and the energy allocation performance of the community microgrid based on the energy shifting \cite{Wang2021Peer}. Similarly, Le et al. \cite{Le2020Tariff} present a load shifting study to find the best schedule load shifting with reduced wind energy curtailment and minimized running costs. When energy storage is introduced to address the fluctuation of power system, Yan et al.  \cite{Yan2020An} propose an allocative method to explore the model between peak-valley difference and hybrid energy storage capacity. However, the classic load shifting methods only consider total loads, instead of the possible demand reductions. To investigate the relationship between the capacity of generation sources and possible demand reductions, Gronier et al. \cite{Gronier2022Iterative} propose an iterative method with demand-side management to determine the sizing selection of photovoltaic panels and solar thermal collectors. {Most of existing studies generally set the incentive price according to individual users' energy data and other more information, which is not easy to obtain or unrealistic. Different from the existing studies, which adopt demand-side management methods, we propose a practical framework to determine the incentive price by indirectly estimating users' willingness to shift based on the public energy data, which helps ensure model scalability and reduces the communication overhead. }
%Few works consider user's willingness to shift with public energy data.

In addition to peak load shifting, renewable energy source and energy storage can also help alleviate the imbalance between supply and demand. However, there is a key challenge to the stochastic characteristic of energy storage and renewable energy resources. Some studies tackle the energy scheduling issues considering the stochastic nature of renewable energy resources. %For instance, 
%, where trading strategy and renewable energy are introduced to reduce energy costs \cite{Daneshvar2020Two, Daryabari2020Stochastic}.
{For instance, a two-stage stochastic programming model equipped with renewable energy resources is presented to handle the uncertainties, which minimizes the imbalance cost online \cite{Daneshvar2020Two}. Likewise, a multi-stage stochastic structure is proposed to integrate flexibility of electrical vehicles into power systems and optimize the operation of electrical vehicles by near real-time optimization with high penetration of renewable energy \cite{Daryabari2020Stochastic}. The utilization of renewable energy and energy storage improves the energy utility \cite{Lak2021Risk, Gomes2019Decision}. In \cite{Lak2021Risk}, two-stage stochastic models are proposed to integrate energy storage systems and wind power producers into frequency regulation, which increase the profit in comparison with other operations. In \cite{Gomes2019Decision}, a strategy based on two-stage stochastic optimization and risk-constrained is addressed for the aggregation of renewable energy and energy storage. %In \cite{Pulendran2019Capacity}, a systematic method for dispatching generation capacities and energy storage in a day-ahead market is proposed and evaluated to solve the problem of resource allocations with minimum cost.
%where incentive prices and energy storage are introduced to reduce energy costs \cite{Deng2014Load,Lakshminarayana2014Cooperation}. In \cite{Deng2014Load}, dual decomposition and stochastic gradient are proposed to address the optimization problem through appropriate scheduling, that is, to shift the peak energy demand by pricing tariffs as incentives. In \cite{Lakshminarayana2014Cooperation}, a stochastic optimization issue is structured to minimize the time-averaged energy cost, and the storage devices are used to integrate renewable energy. The cooperation among prosumers improves the energy utility \cite{Etesami2018Stochastic,Liu2018Energy}. In \cite{Etesami2018Stochastic}, the interaction among prosumers is described as a stochastic game, and a novel distributed algorithm is proposed to achieve optimal returns. In \cite{Liu2018Energy}, a day-ahead scheduling model of an energy sharing provider is built to improve the power profile and increase operating profit through stochastic programming. %
 These studies mainly focus on two-stage stochastic scheduling of electrical load to address the uncertainties of energy storage and renewable energy. We further consider the difference in the characteristics of different energy types and reduce the energy costs by shifting peak load, and adopt two-stage stochastic optimization to achieve the energy storage stability and real-time load balancing. In addition, existing studies rarely consider the improvement of convergence rate, which is of great significance for online energy scheduling. }
 
{ Although some studies \cite{Daneshvar2020Two, Daryabari2020Stochastic, Lohr2021Supervisory} have contributed to the online energy management problem, they rely heavily on explicit predictions of future uncertainties, which is affected by inaccurate choices or models of prediction horizons. Other research attempts to integrate the online operation of energy storage into economical scheduling for ease of implementation. An online energy management is designed to improve the total economy by the coordination scheduling of generation, supercapacitor and battery \cite{Li2021Energy}. A two-stage real-time energy management method is presented by updating the energy set-points of distributed energy resources and adjusting the flexibility injection of energy storage \cite{Kalantar2021Coordinating}.  These online energy management methods are effective to a certain extent by matching the variation tend of renewable energy generation, demand and price, but it is not easy to be precise and adapt to the dynamic changes of the real situation.}

Nowadays, one of the obstacles to multi-energy management in industrial parks is the difference in the timescale of the energy system. For instance, the inelastic loads (ILs), including industrial production demands and refrigerators, need to be satisfied in time, while the elastic loads (ELs), including air conditioners and electric vehicles, can be satisfied later, {and the energy storage devices need stable operation for a long period.} {Considering the temporally-coupled of energy variables, we exploit a two-timescale optimization method to ensure the real-time load balancing on a fast timescale and energy storage stability on a slow timescale by a two-stage relaxation. } In industrial production, the primary and secondary loads are generally not allowed to be interrupted, and part of the tertiary loads can be transferred according to the tightness of the power supply. Considering the pressure of energy supply during peak time, a mechanism is adopted to encourage users to shift partial tertiary loads, 
{where a practical method is proposed to estimate users' willingness to shift IL based on public energy data, instead of obtaining private information of each individual user.}
To solve the issue of energy management, a fast distributed algorithm is proposed to deal with temporally-coupled constraints and ensure real-time realization. % Different from other stochastic mechanisms,  a stochastic optimization problem with two-timescale is constructed to further consider the spatiotemporal changes of load, renewable energy and energy prices. 
Compared with other works, the main contributions of the paper are summarized as follow.
\begin{enumerate}
\item
To satisfy the real-time demand for industrial production, a systematic online energy cost minimization framework for an industrial park is proposed to make full use of the delay-tolerant EL adjustment mechanism, the real-time IL compensation mechanism and the multi source-load-storage coordination mechanism with time-varying generation, demand and price. %A practical method is proposed to estimate users' willingness to shift IL only based on public energy data, instead of knowing energy usage of each individual user.

%, in which EHs provide power and heat for industrial users A multi-energy management framework is presented for an industrial park, where EHs supply electricity and thermal energy to industrial users. This framework fully mobilizes the adjustment mechanism of elastic load (EL), the compensation mechanism of inelastic load (IL) and the coordination mechanism of energy supply, demand and storage. 
% and design distributed algorithm to solve the energy allocation problem for protecting the privacy of users.% incentive users to participate into the energy trading. to solve the energy management problem for protecting the privacy of users, lowering the computational complexity, and realizing faster convergence
\item %The spatiotemporal variation in multi-energy loads, supply and energy prices is considered to constitute a stochastic optimization problem. 
{To achieve two-timescale balances of energy storage on a slow timescale and real-time supply-demand on a fast timescale without knowing statistical information of random variables, a fast distributed algorithm based on dual decomposition and stochastic gradient is proposed to deal with temporally-coupled constraints and get the time-average cost arbitrarily close to the optimal value.}
%To response to multi-energy charge/discharge, intermittent supply of renewable energy, peak load shifting, variable energy prices and demand without knowing the priori knowledge of the underlying random process, a fast distributed algorithm based on two-stage dual decomposition and stochastic gradient is proposed to deal with temporally-coupled constraints and get the time-average cost arbitrarily close to the optimal value.
%In addition to focusing barely on batteries, this paper also considers thermal storage, variable prices, and random multi-energy supply and demand relationships to constitute a two-timescale stochastic optimization problem to % 
%{balance the long-term energy charging and discharging on a slow timescale and satisfy instantaneous loads on a fast timescale while respecting energy constraints.} %consisting of long-term and instantaneous constraints for energy storage and load balance. %, which is independent of statistical characteristics of random events,Both elastic and inelastic loads are considered to construct a two-timescale stochastic network optimization.
%\item According to simulation analyses and results, the proposed algorithm can achieve a better tradeoff among energy cost, energy trading and energy storage. Meanwhile, the effectiveness of the proposed algorithm is verified.
\item 
%The spatiotemporal variation of loads, renewables and energy prices is considered to construct a stochastic network optimization. 
An incentive mechanism is introduced to shift peak load, where a practical method is proposed to estimate users' willingness to shift IL only based on public energy data, instead of knowing energy usage of each individual user.
%In order to further consider the temporal and spatial variation of energy supply and demand, this paper construct a two-timescale optimization to balance the long-term energy charging and discharging on a slow timescale and balance real-time supply-demand on a fast timescale while respecting energy constraints. %
{Through performance analysis and real-data simulation, the feasibility and optimization of the proposed method are proved.} %The fast scheme can ensure real-time coordination of instantaneous scheduling through decentralized implementation.
%\item 
%Different from other stochastic mechanisms,  a stochastic optimization problem with two-timescale is constructed to further consider the spatiotemporal changes of load, renewable energy and energy prices. 

%To reduce the influence of multi-energy coupling, stochastic demand and renewable energy generation, a distributed algorithm independent of statistical characteristics of random events and a fast scheme are proposed. The fast scheme ensures real-time coordination of instantaneous scheduling by a decentralized implementation.
\end{enumerate}

The rest of the paper is structured as follows. Sect. 2 introduces the system model of the multi-energy industrial park. Sect. 3 proposes a fast distributed optimization algorithm based on stochastic gradient and two-stage dual decomposition. %, and its distributed fast realization is presented in Section \uppercase\expandafter{\romannumeral4}. 
Sect. 4 conducts the theoretical analysis of the proposed algorithm. Sect. 5 provides the simulation results and Sect. 6 gives the conclusion and further research. %are given in Section \uppercase\expandafter{6}.\\\改到这儿了

\section{System Model}

\begin{figure}
  \centering
  \includegraphics[width=0.7\hsize]{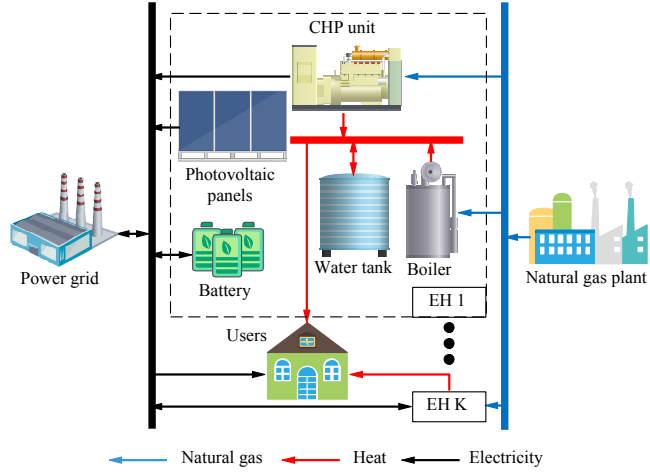}
  \caption{Energy flows of the industrial park}
  \label{fig1}
\end{figure}

%\subsection{Industrial Park}
In this paper, Fig.~\ref{fig1} shows that the industrial park includes EHs and users, where electricity and heat are supplied by power and gas company. %where three types of energies are supplied, i.e., electricity, heat and natural gas, as shown in Fig.~\ref{fig1}. The industrial park is composed of users  and EHs \cite{Yazdani2018Strategic}. 
The EHs are composed of CHP units, batteries, water tanks, photovoltaic panels and boilers.
The CHP units generate heat and electricity with fixed ratio. Batteries and water tanks store electricity and heat, respectively. Photovoltaic panels harvest renewable energy. Boilers generate heat for users. {The hot water in water tanks comes from CHP units and boilers. The heat generated by CHP units and boilers is used to meet the heat demands of users. When the electricity demands of users and the electricity price are high, the CHP units consumes natural gas to generate much electricity and heat. The excess heat will be stored in the water tanks in the form of hot water.}
The park has $\boldsymbol{K}=\{1, 2, ..., K\}$ EHs. The model of EHs $k$ will be given as follows, $k\in \boldsymbol{K}$. %The energy equipments in the next subsection are modeled for EH $k$, $k\in \boldsymbol{K}$. %A list of notations is shown in Table I, where one time slot is one hour to coordinate with the simulation.

\subsection{Energy Hub}
In this subsection, %
the model of EH $k$, which consists of battery $k$, water tank $k$, CHP $k$ and boiler $k$, is given, where one time slot denotes one hour.
\subsubsection{Battery and Water Tank}
%For EH $k$, the amount of electricity stored on a battery and the thermal energy of a hot water tank at time $t$ are denoted as $B_{k}(t)$ and $W_{k}(t)$, respectively. Electric energy and equivalent heat energy are charged in the amount of $C_{ke}(t)$ and $C_{kh}(t)$, and discharged in the amount of  $D_{ke}(t)$ and  $D_{kh}(t)$.
 The battery $k$ and water tank $k$ are modeled as 
\begin{equation}
B_{k}(t+1)=B_{k}(t)+\eta_{cke}C_{ke}(t)-\frac{1}{\eta_{dke}}D_{ke}(t)
\label{A1}
\end{equation}
\begin{equation}
W_{k}(t+1)=W_{k}(t)+\eta_{ckh}C_{kh}(t)-\frac{1}{\eta_{dkh}}D_{kh}(t)
\label{A3}
\end{equation}
%where $\eta_{cke}$ and $\eta_{dke}$ are the charge and discharge efficiencies of the battery. $\eta_{ckh}$ and $\eta_{dkh}$ are the charge and discharge efficiencies of the water tank. 
%In addition, the capacity constraints of the battery and the water tank are
\begin{equation}
B_{k,min}\leq B_{k}(t) \leq B_{k,max}, W_{k,min}\leq W_{k}(t) \leq W_{k,max}
%\label{Bm}
%\end{equation}
%\begin{equation}
%W_{k,min}\leq W_{k}(t) \leq W_{k,max}
\label{Wm}
\end{equation}
%where $B_{k,min}$ and $B_{k,max}$ are the minimal and maximal state of charge (SoC) of the battery, and $B_{k,min}$ is related to the number of charging/discharging cycles \cite{Ke2017Control}. The larger $B_{k,min}$ is, the more charging/discharging cycles can be sustained. Similarly, $W_{k,min}$ and $W_{k,max}$ are the lower and upper bounds of the hot water tank's equivalent thermal energy. The constraints of charging rates of the battery and water tank are $C_{ke,max}$ and $C_{kh,max}$, and the constraints of discharging rates of the battery and water tank are $D_{ke,max}$ and $D_{kh,max}$ during a time slot. 
%where $B_{k,min}$ is related to the number of charge/discharge cycles \cite{Ke2017Control}. The smaller $B_{k,min}$ is, the fewer charge/discharge cycles will be maintained.
%The constraints of charge/dischargee rates of the battery and water tank are
\begin{equation}
0\leq C_{ke}(t) \leq C_{ke,max}, 0\leq D_{ke}(t) \leq D_{ke,max}
\label{Cem}
\end{equation}
\begin{equation}
0\leq C_{kh}(t) \leq C_{kh,max}, 0\leq D_{kh}(t) \leq D_{kh,max}
\label{Chm}
\end{equation}

The battery power $B_{k}(t+1)$ at time slot $t+1$ is equal to the battery power $B_{k}(t)$ at time slot $t$ plus the charge amount $\eta_{cke}C_{ke}(t)$ minus the discharge amount $\frac{1}{\eta_{dke}}D_{ke}(t)$. The thermal energy of water tank $W_{k}(t+1)$ at time slot $t+1$ is equal to the thermal energy of water tank $W_{k}(t)$ at time slot $t$ plus the charge amount $\eta_{ckh}C_{kh}(t)$ minus the discharge amount $\frac{1}{\eta_{dkh}}D_{kh}(t)$. 
%where $B_{k}(t)$, $\eta_{cke}$, $C_{ke}(t)$, $\eta_{dke}$ and $D_{ke}(t)$ denote the stored electricity, charging efficiency, charging rate, discharging efficiency and discharging rate of the battery. $W_{k}(t)$, $\eta_{ckh}$, $C_{kh}(t)$, $\eta_{dkh}$ and $D_{kh}(t)$ denote the stored thermal energy, charging efficiency, charging rate, discharging efficiency and discharging rate of the water tank. %$B_{k,min}$ is related to the number of charge/discharge cycles \cite{Ke2017Control}. The smaller $B_{k,min}$ is, the fewer charge/discharge cycles will be maintained.
\subsubsection{CHP and Boiler}
The models of CHP unit $k$ and boiler $k$ are expressed by
%\begin{subequations}
\begin{equation}
E_{kCHP}(t)=\eta_{kpg}G_{kCHP}(t), H_{kCHP}(t)=\eta_{khg}G_{kCHP}(t)
\label{CHP1}
\end{equation}
\begin{equation}
0\leq E_{kCHP}(t) \leq E_{kCHP,max}, 0\leq H_{kCHP}(t) \leq H_{kCHP,max}
\label{CHP2}
\end{equation}
%\begin{equation}
%H_{kCHP}(t)=\eta_{khg}G_{kCHP}(t)
%\end{equation}
%\label{CHP1}
%\end{subequations}
%where $E_{kCHP}(t)$, $H_{kCHP}(t)$ and $G_{kCHP}(t)$ are the power generation, calorific value and natural gas consumption of the CHP unit at time slot $t$, respectively. $\eta_{kpg}$ and $\eta_{khg}$ are the conversion efficiencies of the CHP unit from natural gas to electricity and heat, respectively. 
%There are maximum electricity $E_{kCHP,max}$ and heat $H_{kCHP,max}$ generation during a time slot. Hence, 
%The generation limits of electricity and heat are:
%\begin{subequations}
%\begin{equation}
%0\leq E_{kCHP}(t) \leq E_{kCHP,max}, 0\leq H_{kCHP}(t) \leq H_{kCHP,max}
%\label{CHP2}
%\end{equation}
%\begin{equation}
%0\leq H_{kCHP}(t) \leq H_{kCHP,max}
%\end{equation}
%\label{CHP2}
%\end{subequations}
%\subsubsection{Boiler}
%A boiler generates heat by consuming natural gas, which is 
\begin{equation}
H_{kb}(t)=\eta_{kbg}G_{kb}(t)
\label{bo1}
\end{equation}
\begin{equation}
0\leq H_{kb}(t) \leq H_{kb,max}
\label{bo2}
\end{equation}

CHP $k$ generates electricity $E_{kCHP}(t)$ and thermal energy $H_{kCHP}(t)$ by utilizing gas $G_{kCHP}(t)$. Boiler $k$ generates thermal energy $H_{kb}(t)$ by utilizing gas $G_{kb}(t)$. The corresponding conversion efficiencies are $\eta_{kpg}$, $\eta_{khg}$ and $\eta_{kbg}$.
%where $E_{kCHP}(t)$, $\eta_{kpg}$, $G_{kCHP}(t)$, $H_{kCHP}(t)$ and $\eta_{khg}$ are the electricity generation, electricity generation efficiency, natural gas consumption, heat generation and heat generation efficiency of the CHP unit at time slot $t$, respectively. where $H_{kb}(t)$, $\eta_{kbg}$ and $G_{kb}(t)$ are the heat generation, heat generation efficiency and natural gas consumption of the boiler, respectively. %$\eta_{kbg}$ is the conversion efficiency of the boiler from natural gas to heat. %$H_{kb,max}$ is the maximum heat generation of the boiler during a time slot. 

\subsection{Energy Trading with the Electricity and Gas Companies}
The park purchases electricity $E(t)$ and gas $G(t)$ from utility companies to supply demand, respectively. When there is extra renewable energy, the excess electricity $E_{o}(t)$ can be sold to the company. %There are maximum constraints of trading energy with public utility companies during a time slot.
%\begin{subequations}
\begin{equation}
0\leq E(t) \leq E_{max},  0\leq G(t) \leq G_{max} , 0\leq E_{o}(t) \leq E_{o,max}
\label{eo2}
\end{equation}
%\begin{equation}
%0\leq G(t) \leq G_{max} 
%\end{equation}
%\begin{equation}
%0\leq E_{o}(t) \leq E_{o,max}
%\end{equation}
%\label{bo2}
%\end{subequations}

\subsection{Energy Load}
%The total available energy for industrial users depends on the energy flows of EHs and utility companies. 
The total available electricity $E_{tot}(t)$, heat $H_{tot}(t)$ and gas $G_{tot}(t)$ for industrial users is denoted as %total available energy can be denoted as:
\begin{subequations}
\begin{equation}
E_{tot}(t)=\sum_{k\in \boldsymbol{K}}E_{kCHP}(t)+D_{ke}(t)-C_{ke}(t)+R_{k}(t)+E(t)-E_{o}(t)
\label{11a}
\end{equation}
\begin{equation}
 G_{tot}(t)=G(t)-\sum_{k\in \boldsymbol{K}}[G_{kCHP}(t)+G_{kb}(t)]
 \label{11b}
 \end{equation}
\begin{equation}
H_{tot}(t)=\sum_{k\in \boldsymbol{K}}H_{kCHP}(t)+H_{kb}(t)+D_{kh}(t)-C_{kh}(t)
\label{11c}
\end{equation}
%\begin{equation}
%E_{k}(t)=E_{kCHP}(t)+D_{ke}(t)-C_{ke}(t)+R_{k}(t)
%\label{11d}
%\end{equation}
%\begin{equation}
%H_{k}(t)=H_{kCHP}(t)+H_{kb}(t)+D_{kh}(t)-C_{kh}(t)
%\label{11e}
%\end{equation}
\end{subequations}
%\\

The total electrical load is supplied by CHP units, batteries, renewable energy $R_{k}(t)$ and the power company. The total gas load is supplied by gas company. The total heat load is supplied by CHP units, boilers and water tanks.
%where (\ref{11a}), (\ref{11b}) and (\ref{11c}) denote the balance of electricity, gas and heat, and (\ref{11d}) and (\ref{11e}) denote the electricity and heat provided by EH $k$. $E_{tot}(t)$, $G_{tot}(t)$ and $H_{tot}(t)$ are the total available electricity, gas and heat, respectively. $E_{k}(t)$ and $H_{k}(t)$ are the electricity and heat generation of EH $k$. $R_{k}(t)$ is the renewable energy generation of EH $k$. For simplicity, heat and gas loads are considered as EL. 

The electrical loads include IL and EL. The IL, such as industrial production demands, refrigerators and illuminations, will not change easily over time. The EL, such as air conditioners, electric vehicles and gas water heaters, can be flexibly arranged. The number of IL users is $\boldsymbol{I}=\{1, 2, ..., I\}$ and the number of EL types is $\boldsymbol{Q}=\{1, 2, ..., Q\}$.
The available energy of EH $k$ is expressed by:
\begin{equation}
0\leq x_{k}(t)\leq x_{k,max}
\label{eqdt1}
\end{equation}
\begin{equation}
x_{k}(t)=\sum_{i \in \boldsymbol{I}}x_{ki}(t)+\sum_{q\in \boldsymbol{Q}}x_{kq}(t)
\label{eqdt}
\end{equation}
where $x_{k}(t)$ denotes the energy generation of EH $k$ with energy $x \in \boldsymbol X$, %.  For example, when there are high energy demands or power outages,
{ and $\boldsymbol X = \{E, H\}$, i.e., a set of electricity and heat.  }$x_{ki}(t)$ denote the IL of user $i$ supplied by EH $k$. $x_{kq}(t)$ denote the EL $q$ of users supplied by EH $k$. 

\subsection{Incentive Mechanism}
Some electrical IL, such as illuminations and inefficient tertiary industrial demands, can be partially shifted when necessary, the industrial park will offer incentive price $p(t')$ to shift  electrical IL of user $i$ by an amount ${X}_{ir}(t'-t)=r_i(p(t'),(t'-t))  {X}_i(t)$ from $t$ to $t'$ without compromising basic demand, where $r_i \leq \eta$ denotes the shifting probability of original electrical IL ${X}_i(t)$ of user $i$. The electrical IL of user $i$ shifted from $t$ is ${X}_{ir}(t)=\sum_{t' \neq t}X_{ir}(t'-t)=\sum_{t' \neq t}r_i(p(t'),(t'-t))  {X}_i(t)$. %${X}_{ir}(t)\leq\eta{X}_i(t)$, where $\eta$ is the ratio of maximum load reduction. %, and ${X}_{i}(t)$ is the original electricity IL of user $i$. 
% The demand shifted from $t'$ to $t$ can be denoted by
%\begin{equation}
%\sum_{i \in \boldsymbol{I}}{X}_i(t')r_i(p(t), (t-t'))
%\label{stt}
%\end{equation}
%where $(t-t')$ denotes that users shift demand from $t'$ to $t$. The total demand shifted to time slot $t$ can be denoted as
%\begin{equation}
%\sum_{t'\neq t}\sum_{i \in \boldsymbol{I}}{X}_{i}(t')r_i(p(t), (t-t'))
%\label{stt}
%\end{equation}

%Thus, the whole cost of offering incentive is denoted as
%\begin{equation}
%\sum_{t=0}^{T-1}p(t')\sum_{t'\neq t}\sum_{i \in \boldsymbol{I}}{X}_{i}(t)r_i(p(t), (t-t'))
%\label{coi}
%\end{equation}

Thus, the cost of offering incentive for the shifted electrical IL of user $i$ from $t$ is denoted as
\begin{equation}
\sum_{t'\neq t}p(t'){X}_{ir}(t'-t)
\label{coi}
\end{equation}

To estimate the shifting functions, a form of given function with adjustable parameters is introduced \cite{Carlee2012Optimized}. {The shifting function $r$ is regarded to exponentially decrease in time and increase in incentive amount:}
\begin{equation}
r_{\alpha_i}(p,t)=C_{\alpha_i}\frac{p}{(t+1)^{\alpha_i}}
\label{rf}
\end{equation}
where $C_{\alpha_i}$ is a constant which depends on $\alpha_i$. The parameter $\alpha_i$ denotes the willingness of user $i$ to shift IL. The higher $\alpha_i$, the shorter the time user $i$ is willing to shift its demand. {The shifting function can be an approximation of the nonlinear mode of user behavior.}

{To estimate $\alpha_i$ for each user, all $\alpha_i$ at a time slot are integrated in one integral shifting function, which sums the shifting functions of all users at that time slot, weighted based on the percentage of electricity consumed by each user.} The integral shifting function at time slot $t$ is denoted as:
\begin{equation}
R(p,t)=\sum_{i \in \boldsymbol{I}}\beta_{i}r_{i}(p,t)
\label{irf}
\end{equation}
where $\beta_{i}$ is the percentage of electricity consumed by user $i$. The amount of electricity demand shifted from $t$ to $t'$ is:
\begin{equation}
A_{t,t'}=X_{IL}(t)R(p(t'), (t'-t))
\label{aes}
\end{equation}
where %$\eta$ is the ratio of maximum load shifting, and 
${X}_{IL}(t)$ is the original electrical IL of all users at time slot $t$.

The difference $J_t$ between the original demand and the demand after offering incentive at time slot $t$, which can be obtained directly from historical data in large quantities, is denoted as:
\begin{equation}
J_t=\sum_{t' \neq t} A_{t',t}-A_{t,t'}
\label{doo}
\end{equation}

Since $t=0,1,...,T-1$, each $J_t$ is denoted as a linear function of multiple $A_{t',t}$. Based on the values from $J_0$ to $J_{T-1}$, these linear expressions can be solved for $A_{t,t'}$. %There is $\frac{T(T-1)}{2}$ $A_{t',t}$ in the equations.  (i.e., $\frac{T(T-1)}{2}$)
%The least-squares or learning methods can be used to calculate $\alpha_i$ and $\beta_i$ to obtain the shifting functions and $A_{t,t'}$ offline. 

According to $A_{t,t'}=X_{IL}(t)\sum_{i \in \boldsymbol{I}}\beta_i C_{\alpha_i}\frac{p(t')}{(t'+1)^{\alpha_i}}$, we assume that $Y(t)=\sum_{i \in \boldsymbol{I}}\beta_i C_{\alpha_i}\frac{p(t)}{(t+1)^{\alpha_i}}$. The error sum of squares is denoted as 
\begin{equation}
\nonumber
\begin{aligned}
&E_Y=\sum_{t' \neq t}(Y(t')-\sum_{i \in \boldsymbol{I}}\beta_i C_{\alpha_i}\frac{p(t')}{(t'+1)^{\alpha_i}})^2%\\&=\sum_{t' \neq t}(\sum_{i \in \boldsymbol{I}}(Y_i(t')-\beta_i C_{\alpha_i}\frac{p(t')}{(t'+1)^{\alpha_i}}))^2
\label{ls}
\end{aligned}
\end{equation}

 Letting $y_{i}(t)=\frac{\beta_i C_{\alpha_i}}{(t'+1)^{\alpha_i}}$, we have $\ln y_{i}(t)=\ln C_{\alpha_i}+\ln \beta_i -{\alpha_i}\ln (t+1)$.
%\begin{equation}
%E_{ss}=\sum_{t' \neq t}(y_i(t')-\ln C_{\alpha_i}p(t')-\ln \beta_i +{\alpha_i}\ln (t'+1))^2
%\label{lss}
%\end{equation}
In real situation, when $p(t')$ is set, $J_{t}$ and $X_{IL}(t)$ can be obtained from electric meter. Taking the derivative of $E_Y$ with respect to $y_{i}(t')$,
\begin{equation}
%\begin{aligned}
%&\frac{\partial E_s}{\partial \alpha_i}=\sum_{t' \neq t}2(Y(t')-\sum_{i \in \boldsymbol{I}}e^{y_i(t')})(\sum_{i \in \boldsymbol{I}}e^{y_i(t')})(-\ln (t'+1)) \\
 \frac{\partial E_Y}{\partial y_{i}(t')}=\sum_{t' \neq t}2(Y(t')-\sum_{i \in \boldsymbol{I}}y_{i}(t')p(t'))Ip(t')
\label{lsa}
%\end{aligned}
\end{equation}

${y_i(t')}$ can be obtained by setting the derivative (\ref{lsa}) equal to zero. Then taking the derivative of $E_y=\sum_{i\in \boldsymbol{I}}(\ln y_{i}(t)-\ln C_{\alpha_i}-\ln \beta_i +{\alpha_i}\ln (t+1))^2$ with respect to $\alpha_i$ and $\ln \beta_i$, respectively
\begin{equation}
\begin{aligned}
&\frac{\partial E_y}{\partial \alpha_i}=\sum_{i\in \boldsymbol{I}}2(\ln y_{i}(t)-\ln C_{\alpha_i}-\ln \beta_i +{\alpha_i}\ln (t+1))\ln (t+1) \\
& \frac{\partial E_y}{\partial \ln \beta_i}=\sum_{i\in \boldsymbol{I}}-2(\ln y_{i}(t)-\ln C_{\alpha_i}-\ln \beta_i +{\alpha_i}\ln (t+1))
\label{lsay}
\end{aligned}
\end{equation}

{By setting the derivative (\ref{lsay}) equal to zero, we can obtain $\alpha_i$ and $\beta_i$ which give the shifting functions offline.} 
%\begin{equation} and $A_{t,t'}$
%\frac{\partial E_s}{\partial \alpha_i}=\sum_{t' \neq t}2(y_{t,t'}-\sum_{i \in \boldsymbol{I}}X_{i}(t)\beta_i C_{\alpha_i}\frac{p(t')}{(t'+1)^{\alpha_i}})^2
%\label{dab}
%\end{equation}

% User $i$ sets its reduced electricity IL by solving the following problem
%\begin{equation}
%\nonumber
%\max_{0\leq {X}_{ir}(t) \leq \eta X_{i}(t)} p_{i}(t){X}_{ir}(t)-a_{i}{X}^{2}_{ir}(t)
%\end{equation}
%where $a_{i}{X}^{2}_{ir}(t)$ is the unsatisfactory cost, and $a_{i}$ can be obtained by learning from historical data. The reason is that historical data includes statistics of smart grid {\cite{Vapnik2000The}}. Learning other coefficients based on historical data has been also studied by using deep neural networks {\cite{Sun2017Learning}}. $\eta$ is the ratio of maximum load reduction, and ${X}_{i}(t)$ is the original electricity IL of user $i$. 
%When the incentive price satisfies
%\begin{equation}
%0\leq p_{i}(t)\leq 2a_{i}\eta {X}_{i}(t)
%\label{pit}
%\end{equation}
%the optimal solution can be obtained by ${X}_{ir}(t) =p_{i}(t)/2a_{i}$, which means that $p_{i}(t)$ and ${X}_{ir}(t)$ are linearly dependent for user $i$ with a linear coefficient of $2a_{i}$. 

%When incentive price $p_{i}(t)$ is given, the actual electricity IL of user $i$ is expressed as
%\begin{equation}
%\sum_{k\in \boldsymbol{K}_{i}}{x}_{ki}(t) =X_{i}(t)-p_{i}(t)/2a_{i}
%\label{ail}
%\end{equation}
%where $\boldsymbol{K}_{i} \subseteq \boldsymbol{K}$ denotes the set of EHs supplied to user $i$. 

\section{Solution Method}
In this paper, a stochastic gradient-based scheduling method is adopted to minimize the time average cost. All random variables are integrated into $\boldsymbol{r}(t)=\{R(t), X(t)\}$, and all optimization variables are integrated into $\boldsymbol{M}(t)$=\{$x_{ki}(t)$, $x_{kq}(t)$, $x_{k}(t)$, $D_{ke}(t)$, $C_{ke}(t)$, $D_{kh}(t)$, $C_{kh}(t)$, $E_{o}(t)$, $E(t)$, $G(t)$, $p(t)$\}. The total cost of the park at time slot $t$ is expressed as 
\begin{equation}
\begin{aligned}
%\phi(t) &= E(t)p_{e}(t)+G(t)p_{g}(t)-E_{o}(t)p_{o}(t)\\
%&+\sum_{i\in \boldsymbol{I}} [p_{i}(t){X}_{ir}(t)-U_{i}(t)]-\sum_{k\in \boldsymbol{K}}\sum_{q\in \boldsymbol{Q}}U_{kq}(t) \\
\phi(t) &= E(t)p_{e}(t)+G(t)p_{g}(t)-E_{o}(t)p_{o}(t)\\
&+\sum_{i \in \boldsymbol{I}}(\sum_{t'\neq t}p(t'){X}_{ir}(t'-t)-U_{i}(t))-\sum_{k\in \boldsymbol{K}}\sum_{q\in \boldsymbol{Q}}U_{kq}(t)
\end{aligned}
\label{eq3011}
\end{equation}
where $U_{i}(t)=a_{ii}(X_{i}(t)-{X}_{ir}(t))^{2}+b_{ii}(X_{i}(t)-{X}_{ir}(t))$ and $U_{kq}(t)=a_{kq}x_{kq}^{2}(t)+b_{kq}x_{kq}(t)$. $a_{ii}$, $b_{ii}$, $a_{kq}$ and $b_{kq}$ are the corresponding utility coefficients. $p_{e}(t)$ and $p_{g}(t)$ denote the electricity price and gas price of the companies, respectively. $p_{o}(t)$ denotes the electricity price sold to the power company. $U_{i}(t)$ and $U_{kq}(t)$ denote the satisfaction income of IL of user $i$ and EL $q$ provided by EH $k$.

The energy management problem of the park is to find a method to minimize the time average energy cost:
\begin{align}
\min_{\boldsymbol{M}(t)}\lim_{T\rightarrow\infty}\frac{1}{T}\sum_{t=0}^{T-1}\mathbb{E}\{\phi(t)\}
\label{eq301}
\end{align}
\begin{equation}
\nonumber
\text{s.t. } (\ref{A1}) - (\ref{eqdt})
\end{equation}
where the expectation is taken for all uncertain variables.

\begin{figure}
  \centering
  \includegraphics[width=0.8\hsize]{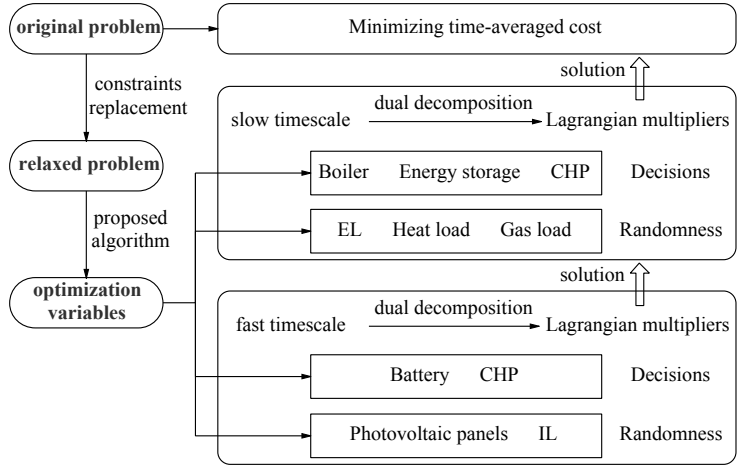}
  \caption{Problem-solving process }
  \label{fig11}
\end{figure}

Some methods in the following subsections are adopted to make the optimization problem easier to solve. Fig.~\ref{fig11} shows the problem-solving process.

\subsection{Constraints Replacement}
The optimization variables are coupled in the battery dynamics (\ref{A1}) and water tank dynamics (\ref{A3}), and cannot be obtained directly. Moreover, considering that it is causal for the knowledge of the random variable $\boldsymbol{r}(t)$, the optimization problem of cross-time coupling is usually difficult to tackle. To handle this issue, we replace the temporally-coupled constraints (\ref{A1}) and (\ref{A3}) with the time-average constraints. %Then, dual decomposition is adopted to separate the problem across time, which is shown in the following section. 

According to (\ref{A1})-(\ref{Wm}), the average energy charge and discharge can be denoted by

\begin{subequations}
\begin{equation}
\lim_{T\rightarrow\infty}\frac{1}{T}\sum_{t=0}^{T-1}\mathbb{E}\{C_{ke}(t)\}=\lim_{T\rightarrow\infty}\frac{1}{T}\sum_{t=0}^{T-1}\mathbb{E}\{D_{ke}(t)\}
\end{equation}
\begin{equation}
\lim_{T\rightarrow\infty}\frac{1}{T}\sum_{t=0}^{T-1}\mathbb{E}\{C_{kh}(t)\} =\lim_{T\rightarrow\infty}\frac{1}{T}\sum_{t=0}^{T-1}\mathbb{E}\{D_{kh}(t)\}
\end{equation}
\label{sto1}
\end{subequations}
where (\ref{sto1}) can guarantee that the charged energy is equal to the discharged energy for a long period. Based on (\ref{sto1}),  the optimization problem (\ref{eq301}) can be relaxed as
\begin{equation}
\begin{aligned}
\min_{\boldsymbol{M}(t)}\lim_{T\rightarrow\infty}\frac{1}{T}\sum_{t=0}^{T-1}\mathbb{E}\{\phi(t)\}
\end{aligned}
\label{eq31}
\end{equation}
\begin{equation}
\nonumber
\text{s.t. } (\ref{Cem}) - (\ref{eqdt}), (\ref{sto1}) 
\end{equation}

Compared with (\ref{eq301}) , time-coupling constraints (\ref{A1}) and (\ref{A3}) are superseded by (\ref{sto1}). %Since (\ref{eq31}) is a relaxed version of (\ref{eq301}), it satisfied that $\Phi_{2}^{*}\leq \Phi_{1}^{*}$, where $\Phi_{1}^{*}$ and $\Phi_{2}^{*}$ are the optimal solution of cost for the original and relaxed problem, respectively. 
If the stochastic process $\boldsymbol{r}(t)$ is stationary, the time-invariant control strategy $\boldsymbol{M}: \boldsymbol{r}(t) \rightarrow \boldsymbol{M}(t)$ induces the solution $\boldsymbol{M}(t)=\boldsymbol{M}(\boldsymbol{r}(t))$, which can satisfy the constraint condition (\ref{Cem}) - (\ref{eqdt}) and (\ref{sto1}),  and achieve the optimal performance \cite{Neely2010Stochastic}. This means that all expectations of the limiting time-averaged in (\ref{eq31}) produce the same result. Thus, the time-averaged can be removed. To deal with the coupling introduced by the expectation in (\ref{sto1}), the constraints (\ref{sto1}) are dualized and the dual decomposition method is used. After the dualization, the solution can be computed separately across time, which will be illustrated in the following section.

\subsection{Stochastic Energy Optimization} 
First, Lagrangian multiplier method is used to deal with the coupling constraint (\ref{sto1}). The Lagrangian function about (\ref{eq31}) is denote as
\begin{equation}
\begin{aligned}
L(\boldsymbol{M}(t), \boldsymbol{\lambda})&=\mathbb{E}\{\phi(t)\}+\sum_{k\in \boldsymbol{K}}\mathbb{E}[\lambda_{ke}(C_{ke}(t)-D_{ke}(t))]\\&+\sum_{k\in \boldsymbol{K}}\mathbb{E}[\lambda_{kh}(C_{kh}(t)-D_{kh}(t))]
\end{aligned}
\label{micro11}
\end{equation}
 where $\lambda_{ke}$ and $\lambda_{kh}$ are the corresponding Lagrangian multipliers.
 
The corresponding Lagrangian dual function is denoted as
\begin{equation}
\begin{aligned}
\Gamma(\boldsymbol{\lambda})&=\min_{\boldsymbol{M}(t)\in \widetilde {\boldsymbol{M}}(t)}L(\boldsymbol{M}(t), \boldsymbol{\lambda})
\end{aligned}
\label{micro12}
\end{equation}
where  $\widetilde {\boldsymbol{M}}(t)$ denotes the solution under constraints (\ref{Cem}) - (\ref{eqdt}). The corresponding dual problem is expressed by 
\begin{equation}
\begin{aligned}
\max_{\boldsymbol{\lambda}}\Gamma(\boldsymbol{\lambda})
\end{aligned}
\label{micro13}
\end{equation}

A gradient algorithm is introduced to obtain the optimal solution $\boldsymbol{\lambda}^{*}$ of problem (\ref{micro13}). The Lagrangian multipliers $\boldsymbol{\lambda}(t+1)$ at time slot $t+1$ are expressed by 

\begin{subequations}
\begin{equation}
\lambda_{ke}(t+1)=\lambda_{ke}(t)+\rho (C_{ke}(t)-D_{ke}(t))
\end{equation}
\begin{equation}
\lambda_{kh}(t+1)=\lambda_{kh}(t)+\rho (C_{kh}(t)-D_{kh}(t))
\end{equation}
\label{micro16}
\end{subequations}
\\where $C_{ke}(t)$, $D_{ke}(t)$, $C_{kh}(t)$ and $D_{kh}(t)$ can be acquired by solving 
\begin{equation}
\begin{aligned}
 \Phi^*&=\min_{M(t)} \Phi(t)=\min_{M(t)} \phi(t)+\sum_{k\in \boldsymbol{K}}[\lambda_{ke}(t)(C_{ke}(t)\\&-D_{ke}(t))+\lambda_{kh}(t)(C_{kh}(t)-D_{kh}(t))]
\end{aligned}
\label{micro17}
\end{equation}
\begin{equation}
\nonumber
\text{s.t. } (\ref{Cem}) - (\ref{eqdt})
\end{equation}

%Due to the dual decomposition of variables over time, stochastic iteration is feasible. 
The stochastic iteration in (\ref{micro16}) and (\ref{micro17}) has two advantages. First, the solution obtained by (\ref{micro17}) approaches the solution to (\ref{eq31}). Second, when initialized correctly, the solution of (\ref{micro17}) can be feasible for original problem (\ref{eq301}). These advantages will be elaborated in Section 4.

\subsection{Fast Distributed Implementation}
Although the original problem (\ref{eq301}) is separated across time, it still requires a centralized method to solve the convex problem (\ref{micro17}), which is challenging in the presence of a large number of variables. Therefore, it is necessary to obtain the optimal solution in a distributed algorithm \cite{Zhang2017Distributed}, which aims to improve robustness and reduce computational complexity.

In this way, the dual decomposition method is introduced again to deal with the coupling between EL and IL for constraint (\ref{eqdt}) in (\ref{micro17}). {The IL should be satisfied immediately and the decision is real-time \cite{Guo2012Optimal}, so the fast distributed algorithm needs to run some iterations on mini-slots at each time slot. That is, the proposed algorithm runs on two timescales.}  The corresponding Lagrangian function on a fast timescale is expressed by 
\begin{equation}
\begin{aligned}
\overline{L}(\boldsymbol{M}, \boldsymbol{\tau}): =\Phi(t)+\sum_{k\in \boldsymbol{K}}\tau_{k}(\sum_{i\in \boldsymbol{I}}x_{ki}+\sum_{q\in \boldsymbol{Q}}x_{kq}-x_{k})
\end{aligned}
\label{micro18}
\end{equation}
where $\boldsymbol{\tau}=\{\tau_{1}, ..., \tau_{K}\}$ denotes the set of the corresponding Lagrangian multipliers, and the dual variable $\boldsymbol{\lambda}(t)$ in $\Phi(t)$ is updated on a slow timescale, i.e., $\boldsymbol{\lambda}(t)$ can be seen as constants to solve optimization problem (\ref{micro17}). The corresponding dual function is expressed as 
\begin{equation}
\overline{\Gamma}(\boldsymbol{\tau}) :=\min_{\boldsymbol{M}\in \overline{\boldsymbol{M}}}\overline{L}(\boldsymbol{M}, \boldsymbol{\tau})
\label{micro19}
\end{equation}
where $\overline{\boldsymbol{M}}$ denotes the feasible solution under constraints (\ref{Cem}) - (\ref{eqdt}). The corresponding dual problem is expressed by 
\begin{equation}
\max_{\boldsymbol{\tau}} \overline{\Gamma}(\boldsymbol{\tau})
\label{micro20}
\end{equation}

Unlike the dual problem (\ref{micro13}), which implements stochastic estimation, (\ref{micro20}) aims to schedule energy in a  distributed manner for each EH and user. Next, two gradient-based methods will be used to solve the issue (\ref{micro20}).

We first introduce the conventional gradient algorithm to find Lagrangian multipliers $\boldsymbol{\tau}$, and the iteration for $\boldsymbol{\tau}$ is expressed by 
\begin{equation}
\boldsymbol{\tau}(n+1)=\boldsymbol{\tau}(n)+\sigma\nabla \overline{\Gamma}(\boldsymbol{\tau}(n))
\label{micro21}
\end{equation}
where $n$ denotes the index of the mini-slot, and $\sigma$ denotes the stepsize. The gradient of ${\tau}_k(n)$ can be expressed as
\begin{equation}
\nabla \overline{\Gamma}(\tau_{k}(n))=\sum_{i\in \boldsymbol{I}}x_{ki}(n)+\sum_{q\in \boldsymbol{Q}}x_{kq}(n)-x_{k}(n)
\label{micro22}
\end{equation}

Although conventional gradient algorithm has been universally used, it does not take advantage of the properties of the problem, i.e., the differentiability of the function and the Lipschitz continuity of the gradient, and its convergence rate is slow. %
{To solve the problem online, a scheme is proposed to realize faster convergence than the conventional gradient algorithm based on a fast iterative algorithm \cite{Beck2009A}. }

Unlike the conventional gradient algorithm of (\ref{micro21}), which only depends on the current iteration, the proposed fast method uses the memory of the previous iteration and constructs $\overline{\boldsymbol{\tau}}(n)$ by utilizing $\boldsymbol{\tau}(n)$ and $\boldsymbol{\tau}(n-1)$, which can be expressed as
\begin{equation}
\overline{\boldsymbol{\tau}}(n)=(1-\epsilon)\boldsymbol{\tau}(n)+\epsilon \boldsymbol{\tau}(n-1)
\label{micro25}
\end{equation}
where $\epsilon=(1-\theta_{\tau}(n-1))/\theta_{\tau}(n)$, and $\theta_{\tau}(n)$ is updated as
\begin{equation}
\theta_{\tau}(n)=(1+\sqrt{1+4\theta^{2}_{\tau}(n-1)})/2
\label{micro26}
\end{equation}

$\boldsymbol{\tau}(n)$ can be denoted using a gradient iteration about $\overline{\boldsymbol{\tau}}(n)$
\begin{equation}
\tau_{k}(n+1)=\overline{\tau}_{k}(n)+\sigma(\sum_{i\in \boldsymbol{I}}x_{ki}(n)+\sum_{q\in \boldsymbol{Q}}x_{kq}(n)-x_{k}(n))
\label{micro27}
\end{equation}

According to ($\ref{micro19}$), the optimal energy scheduling solution $\boldsymbol{M}(n)$ is obtained by solving %=\{$E(n)$, $G(n)$, $E_{o}(n)$, $x_{k}(n)$, $C_{ke}(n)$, $D_{ke}(n)$, $C_{kh}(n)$, $D_{kh}(n)$\} by solving 
\begin{equation}
\begin{aligned}
\min_{\boldsymbol{M}(n)} & \, E(n)p_{e}(n)+G(n)p_{g}(n)-E_{o}(n)p_{o}(n)+\lambda_{ke}(n)(C_{ke}(n)-D_{ke}(n))\\&+\lambda_{kh}(n)(C_{kh}(n)-D_{kh}(n))-\tau_{k}(n)x_{k}(n)+\sum_{k\in \boldsymbol{K}_{i}}\tau_{k}(n)x_{ki}(n)\\&+\sum_{i\in \boldsymbol{I}}[\sum_{n'\neq n}p(n'){X}_{ir}(n'-n)-U_{i}(n)]+\tau_{k}(n)x_{kq}(n)-U_{kq}(n)
\end{aligned}
\label{micro23}
\end{equation}
\begin{equation}
\nonumber
\text{s.t. } (\ref{Cem}) - (\ref{eqdt1}) %, (\ref{pit}) - (\ref{ail})
\end{equation}
%and an EL schedule $x_{kq}(n)$ by solving 
%\begin{equation}
%\begin{aligned}
%\min_{x_{kq}(n)} \tau_{k}(n)x_{kq}(n)-U_{kq}(n)
%\end{aligned}
%\label{micro24}
%\end{equation}
%and obtain \{$x_{ki}(n)$, $p_{i}(n)$\} by solving 
%\begin{equation}
%\min_{x_{ki}(n), X_{ir}(n)} \sum_{k\in \boldsymbol{K}_{i}}\tau_{k}(n)x_{ki}(n)+p_{i}(n){X}_{ir}(n)-U_{i}(n)
%\label{micro241}
%\end{equation}
%\begin{equation}
%\nonumber
%\text{s.t. } (\ref{pit}) - (\ref{ail})
%\end{equation}

{According to the expression of $U_{i}(n)$, }
\begin{equation}
\begin{aligned}
&\sum_{i\in \boldsymbol{I}}[\sum_{n'\neq n}p(n'){X}_{ir}(n'-n)-U_{i}(n)]=\sum_{i\in \boldsymbol{I}}[\sum_{n'\neq n}p(n')X_{i}(n)r_{i}(p(n'),(n'-n))\\&-a_{ii}X^2_{i}(n)(1-\sum_{n'\neq n}r_{i}(p(n'),(n'-n)))^2+b_{ii}X_{i}(n)(1-\sum_{n'\neq n}r_{i}(p(n'),(n'-n)))]
\end{aligned}
\label{td0}
\end{equation}

{For simplicity, $r_{i}(p(n'),(n'-n))$ is written as $r_{i}$. According to (\ref{rf}), $r_i$ and $p(n')$ are linearly related, and $r_i=p(n')\frac{\partial r_{i}}{\partial p(n')}$. Taking the derivative of (\ref{micro23}) with respect to $p(n')$ yields}
\begin{equation}
\begin{aligned}
%&\frac{\partial [\sum_{n'\neq n}p(n'){X}_{ir}(n'-n)-U_{i}(n)]}{\partial p(n')}\\=
&\sum_{i\in \boldsymbol{I}}[2p(n')\frac{\partial r_{i}}{\partial p(n')}X_{i}(n)+2a_{ii}X^2_{i}(n)(1-p(n')\frac{\partial r_{i}}{\partial p(n')})\frac{\partial r_{i}}{\partial p(n')}\\&-b_{ii}X_{i}(n)\frac{\partial r_{i}}{\partial p(n')}]=0
\end{aligned}
\label{td1}
\end{equation}

{$p(n')$ can be denoted by}%$r_{\alpha_i}(p,t)=C_{\alpha_i}\frac{p}{(t+1)^{\alpha_i}}$
\begin{equation}
\nonumber
\begin{aligned}
&p(n')=\frac{\sum_{i\in \boldsymbol{I}}(X_{i}(n)\frac{\partial r_{i}}{\partial p(n')})(2a_{ii}X_{i}(n)-b_{ii})}{\sum_{i\in \boldsymbol{I}}(X_{i}(n)\frac{\partial r_{i}}{\partial p(n')})(2a_{ii}X_{i}(n)\frac{\partial r_{i}}{\partial p(n')}-2)}
\end{aligned}
\label{td1}
\end{equation}
%and $p(n')$ can be obtained. %by taking the derivative of $p(n)\sum_{t'\neq t}{X}_{i}(n')r_i(p(n), (n'-n))-U_{i}(n)=p(n)\sum_{t'\neq t}{X}_{i}(n')r_i(p(n), (n'-n))-a_{ii}(X_{i}(t)-{X}_{ir}(t))^{2}+b_{ii}(X_{i}(t)-{X}_{ir}(t))$ with respect to $p(n)$ based on the incentive mechanism in Section II. 

Considering that (\ref{micro17}) satisfies the Slater condition and  is convex, the solution obtained by (\ref{micro20}) is feasible for the original problem (\ref{micro17}). When $\rho$ is properly chosen, the iterations (\ref{micro21}) and (\ref{micro27}) will converge to the vicinity of $\boldsymbol{\tau}^{*}$, and the obtained solution can be made arbitrarily close to the optimal value \cite{Bertsekas2009Convex}. %The iterations will be stopped when a pre-specified tolerance is met.  \cite{Beck2009A}, which incurs low complexity $\mathcal{O}(I(I^2+K^3))$

{The proposed fast method is shown in Algorithm 1, which incurs low complexity. Updating (\ref{micro16}) only requires complexity $\mathcal{O}(K)$, and the complexity of worst-case for solving (\ref{micro23}) is $\mathcal{O}(I(I^2+K^3))$.} {By combining the two consecutive iterations, the constructed iteration $\overline{\tau}(n)$ achieves faster convergence by reducing the undesired fluctuation of the gradient ascent method without compromising accuracy. The convergence of Algorithm 1 needs to meet two conditions: 1) the dual function $\overline{\Gamma}(\boldsymbol{\tau})$ is differentiable; 2) the gradient $\nabla \overline{\Gamma}$ is Lipschitz continuous.  In practice, marginal costs are usually monotonic, which guarantee that the cost function $\phi(t)$ is strongly convex about $\boldsymbol{M}(t)$. For a given $\boldsymbol{\tau}$, the Lagrangian function (30) has a unique minimizer. Therefore, the dual function $\overline{\Gamma}(\boldsymbol{\tau})$ is differentiable and the gradient $\nabla \overline{\Gamma}$ is Lipschitz continuous.}

The detailed proof of condition 1) can be found in \cite{Low1999Optimization}, and the detailed proof of condition 2) can be found in \cite{Beck2014An}. 

\begin{algorithm}[h]
\floatname{algorithm}{Algorithm}
\renewcommand{\algorithmicrequire}{\textbf{Communication}}
\renewcommand{\algorithmicensure}{\textbf{Contribution}}
\footnotesize
\caption{Fast Method}
\label{alg1}
\begin{algorithmic}[1]
  \State \textbf{Initialize:} $\boldsymbol{\tau}(0)$, $\boldsymbol{\tau}(1)$, $\theta_{\tau}(0)$ and $\sigma$.   
    \For{each mini-slot $n$}
\State Calculate $\theta_{\tau}(n)$ based on (\ref{micro26}).
\State Calculate $\overline{\boldsymbol{\tau}}(n)$ according to (\ref{micro25}).
\State EHs send $\overline{\boldsymbol{\tau}}(n)$ to users.
\State Each user and EH obtain the solution $\boldsymbol{M}(n)$ using $\boldsymbol{\tau}=\overline{\boldsymbol{\tau}}$ according to (\ref{micro23}). % - (\ref{micro241}).
\State Update $\boldsymbol{\tau}(n+1)$ based on (\ref{micro27}).
\EndFor 
% \State \ \textbf{end for} 
\label{code:recentEnd}
\end{algorithmic}
\end{algorithm}

\begin{figure}
  \centering
  \includegraphics[width=.8\hsize]{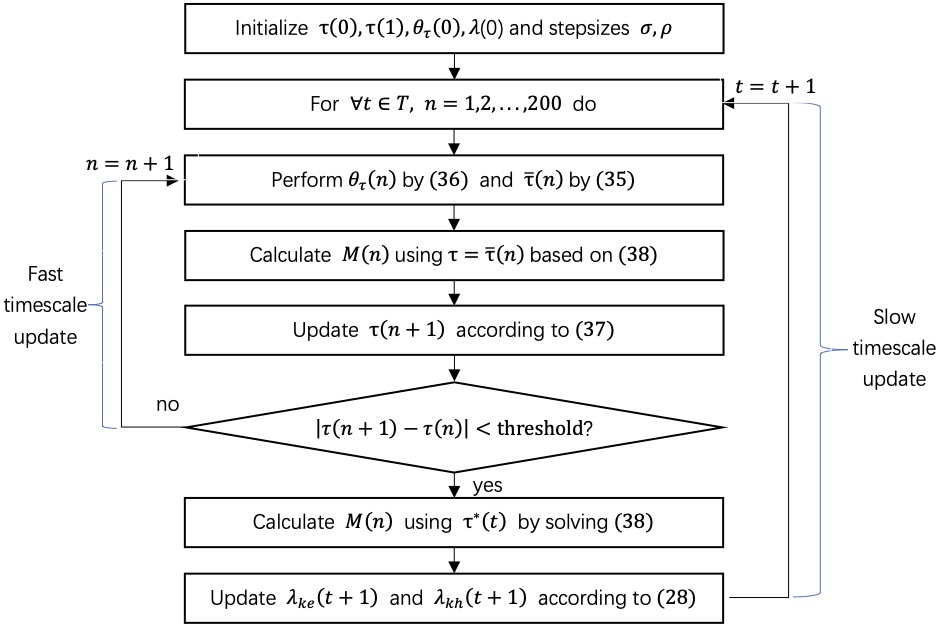}
  \caption{{Flow chart of the proposed algorithm}}
  \label{fig10}
\end{figure}

The proposed fast distributed algorithm, which combines the dual decomposition and the fast method, is shown in Fig.~\ref{fig10}. %Algorithm 2.
%\begin{algorithm}[h]
%\floatname{algorithm}{Algorithm}
%\renewcommand{\algorithmicrequire}{\textbf{Communication}}
%\renewcommand{\algorithmicensure}{\textbf{Contribution}}
%\footnotesize
%\caption{: Distributed Stochastic Gradient Algorithm}
%\label{alg2}
%\begin{algorithmic}[1]
%  \State \textbf{Initialize:} $\boldsymbol{\lambda}(0)$, and stepsize $\rho$.   
%      \For{$t= 1, 2, ...$}
%\State Obtain $\boldsymbol{\tau}^{*}(t)$ according to (\ref{micro21}) or (\ref{micro27}).
%\State User $i$ performs $ x_{ki}(t)$ and ${X}_{ir}(t)$ using $\boldsymbol{\tau}^{*}(t)$ by solving (\ref{micro241}).
%\State Perform $x_{kq}(t)$ using $\boldsymbol{\tau}^{*}(t)$ by solving (\ref{micro24}).
%\State Perform $\boldsymbol{E}_{a}(t)$ using $\boldsymbol{\tau}^{*}(t)$ by solving (\ref{micro23}).
%\State Based on the solution $C_{ke}(t)$, $D_{ke}(t)$, $C_{kh}(t)$ and $D_{kh}(t)$, EH $k$ updates $\lambda_{ke}(t+1)$ and $\lambda_{kh}(t+1)$ according to (\ref{micro16}).      
%\EndFor 
%\label{code:recentEnd2}
%\end{algorithmic}
%\end{algorithm}

%The flow chart of the proposed algrorithm is shown in Fig.~\ref{fig10}. 

\section{Performance Analysis}

To enable the proposed algorithm to generate feasible strategy for (\ref{eq301}), the following properties are given:

\textbf{Lemma 1.} The charge and discharge of the battery satisfy: 1) When $\lambda_{ke}(t)>-{p}_{o}(t)$, $D_{ke}(t)={D}_{ke,max}$ ; 2) When $\lambda_{ke}(t)<-{p}_{e}(t)$, $C_{ke}(t)={C}_{ke,max}$. The equivalent thermal energy of charge and discharge of the water tank satisfy: 1) When $\lambda_{kh}(t)>0$, $D_{kh}(t)={D}_{kh,max}$; 2) When $\lambda_{kh}(t)<0$, $C_{kh}(t)={C}_{kh,max}$. 

For brevity, the proof is omitted. Please refer to Ref. \cite{Zhu2020Energy}.

Lemma 1 states that the Lagrangian multiplier $\lambda_{ke}(t)$ can be regarded as a charge price. When $\lambda_{ke}(t)$ is low enough, the battery will charge as it can. When $\lambda_{ke}(t)$ is high enough, the battery will discharge as it can. According to Lemma 1, the result can be determined as follows:

\textbf{Lemma 2.} If $\rho\geq {\rho_{min}}$, where ${\rho}_{min}=\max_{k}\frac{{p}_{e,max}-{p}_{o,min}}{ {B}_{k,max}-{B}_{k,min}-{D}_{ke,max}-{C}_{ke,max}}$, Lagrangian multipliers $\lambda_{ke}(t)$ and $\lambda_{kh}(t)$ satisfy $-{p}_{e,max}-\rho {D}_{ke,max}\leq \lambda_{ke}(t)\leq \rho {B}_{k,max}-\rho {B}_{k,min}-{p}_{e,max}-\rho {D}_{ke,max}$ and $-\rho {D}_{kh,max}\leq \lambda_{kh}(t)\leq \rho {W}_{k,max}-\rho {W}_{k,min}-\rho {D}_{kh,max}$.

The proof of this step is shown in Appendix A.

According to Lemma 2, when $B_{k}(t)=(\lambda_{ke}(t)+ {p}_{e,max} )/\rho+ {B}_{k,min}+ D_{ke,max}$ and $W_{k}(t)=\lambda_{kh}(t) /\rho+ {W}_{k,min}+ D_{kh,max}$, $ {B}_{k,min}\leq {B}_{k}\leq {B}_{k,max}$ and ${W}_{k,min}\leq {W}_{k}\leq {W}_{k,max}$, i.e., the capacity constraints of the battery and water tank are always satisfied. 

Based on Lemma 2, the performance of the fast distributed algorithm can be obtained.

\textbf{Theorem 1.} When the random state $\boldsymbol{r}(t)$ is i.i.d, $\lambda_{ke}(0) = \rho {B}_{k}(0)-\rho {B}_{k,min}-{p}_{e}(0)-\rho {D}_{ke,max}$, and $\rho\geq {\rho_{min}}$, the expected time-averaged cost under the fast distributed algorithm satisfies
\begin{align}
\lim_{T\rightarrow\infty}\frac{1}{T}\sum_{t=0}^{T-1}\mathbb{E}\{\phi(t)\}\leq\phi^{*}+\rho F
\end{align}
where $\phi^{*}$ is the optimal cost of (\ref{eq301}), and $F$ is defined by
\begin{equation}
\begin{aligned}
F=\frac{1}{2}(\max({C}_{ke, max},{D}_{ke, max}))^2+\frac{1}{2}(\max(C_{kh,max},D_{kh,max}))^{2}
\end{aligned}
\end{equation}

{\it Proof:} For brevity, the proof is omitted here.  Please refer to Ref. \cite{Zhu2020Energy}.

Theorem 1 shows the gap between the performance of the fast distributed algorithm and the optimal performance. The gap increases with $\rho$. According to Lemma 2, it can be concluded that the smaller ${p}_{e,max}-{p}_{o,min}$, the smaller $\rho$. Similarly, the larger $ B_{k,max}$ in Lemma 2, the smaller $\rho$. Therefore, when ${p}_{e,max}-{p}_{o,min}$ is made close to zero or $ B_{k,max}$ is large, the stepsize $\rho$ will be small enough and the gap will be close to zero.

\section{Case Studies}

We give numerical results based on actual data to evaluate the performance of proposed algorithm .
\subsection{Simulation Setup}
Three factories and two EHs are considered in the industrial park. Each EH has a battery, a hot water tank, a CHP unit and a boiler. Each factory has one kind of electrical EL and IL, gas and heat loads, respectively. For simplicity, the three kinds of electrical EL in the three factories are calculated together and shown as a whole in the simulation figure. %
{The coefficients of the same type of equipment and load are set to be the same. }For example, energy conversion coefficients of two boilers are both  $\eta_{kbg}=85\%$. The same assumption is available for batteries, water tanks, CHP units, ILs and ELs.  The ratio of maximum electrical IL shifting is $\eta=0.15$, and the gas price is $p_{g}(t)=0.4$ \textyen/kWh. %
{Other relevant parameters are shown in Table 1, and the parameter settings are similar to Ref. \cite{Wang2018Incentivizing}. The adjustment of the parameters will not affect the simulation results.}
% The unsatisfactory coefficient is $a_{i}=1$, and the utility coefficients are $a_{ii}=-1$, $b_{ii}=1$, $a_{kq}=-1$ and $b_{kq}=7$, respectively. The parameters of efficiency are $\eta_{cke}=\eta_{dke}=\eta_{ckh}=\eta_{dkh}=98\%$, $\eta_{kpg}=\eta_{khg}=35\%$, respectively. Other parameters are summarized as follows: 
% $p_{g}(t)=0.4$ yuan/kWh, $B_{k,max}=4$MWh, $C_{ke,max}=D_{ke,max}=1$MWh, $W_{k,max}=4$MWh, $C_{kh,max}=D_{kh,max}=1$MWh. %The EL has 2 types. 

%\begin{table}\footnotesize
%\centering
%\caption{Relevant Parameters}
%\begin{tabular}{m{1.4cm}|m{6.2cm}}
%\begin{tabular}{l l l }
%\hline
%$B_{k,max}, W_{k,max}$ & $C_{ke,max}$, $D_{ke,max}$ & $C_{kh,max}$, $D_{kh,max}$ \\
%\hline
%4MWh & 1MWh & 1MWh  \\
%\hline
%\end{tabular}
% \\ [5pt]
%\begin{tabular}
%\hline
%$b_{ii}$ & $a_{ii}$, $a_{kq}$ & $b_{kq}$ & $\eta_{kpg}$ & $\eta_{khg}$ & $\eta_{cke}$, $\eta_{dke}$, $\eta_{ckh}$, $\eta_{dkh}$\\
%\hline
%1 & -1 & 7 & 35\% & 45\% & 98\% \\
%\hline
%\end{tabular}
%\label{TAB1}
%\end{table}

\begin{table}\footnotesize
\centering
\caption{Relevant Parameters}
\begin{tabular}{|m{3cm}m{1cm}|m{3cm}m{1cm}|}
\hline
Parameter &Value & Parameter & Value \\
\hline
$B_{k,max}$,$W_{k,max}$& 4MWh  & $a_{ii}$,$a_{kq}$ &-1\\

$C_{kh,max}$,$D_{kh,max}$ & 1MWh&$b_{ii}$& 1   \\

$C_{ke,max}$,$D_{ke,max}$& 1MWh&$b_{kq}$& 7 \\

$\eta_{kpg}$&35\%  & $\eta_{khg}$& 45\%\\

$\eta_{cke}$,$\eta_{dke}$&98\%& $\eta_{ckh}$,$\eta_{dkh}$&98\%  \\
\hline
\end{tabular}
\label{TAB1}
\end{table}

\begin{figure}
\centering
\begin{minipage}{1\linewidth}
  \centerline{\includegraphics[width=0.7\textwidth]{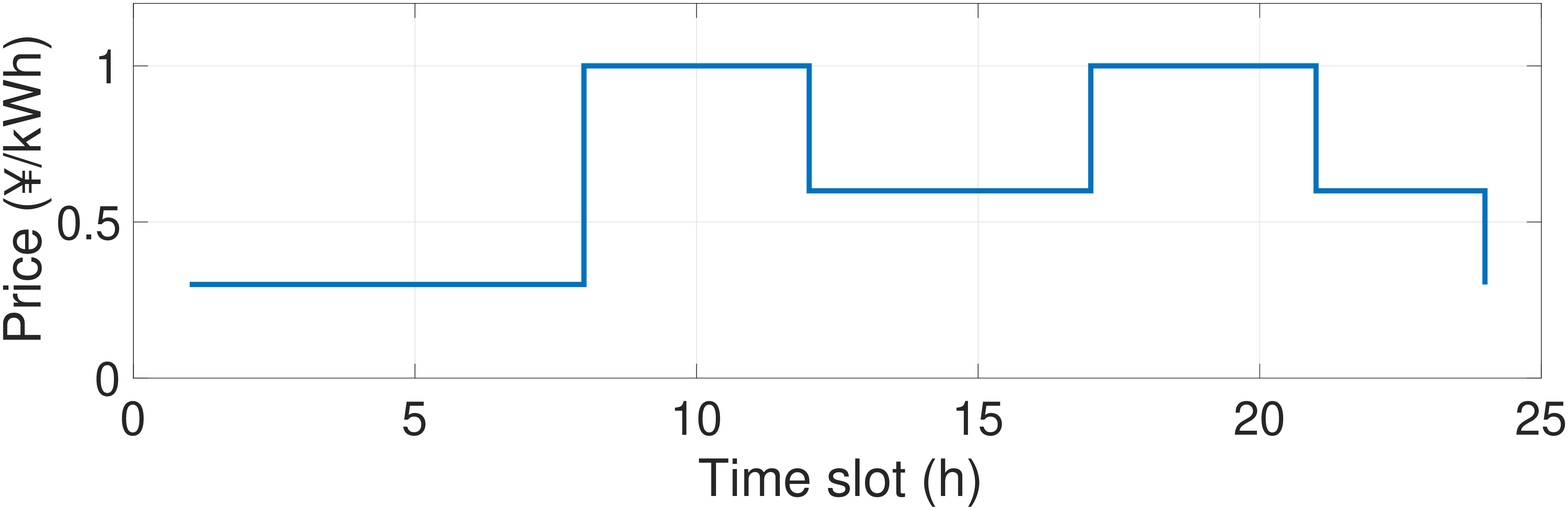}}
  \centerline{\scriptsize{(a) Electricity price of the power company}}
\end{minipage}
%\hspace{-5pt}
\begin{minipage}{1\linewidth}
 \centerline{\includegraphics[width=0.7\textwidth]{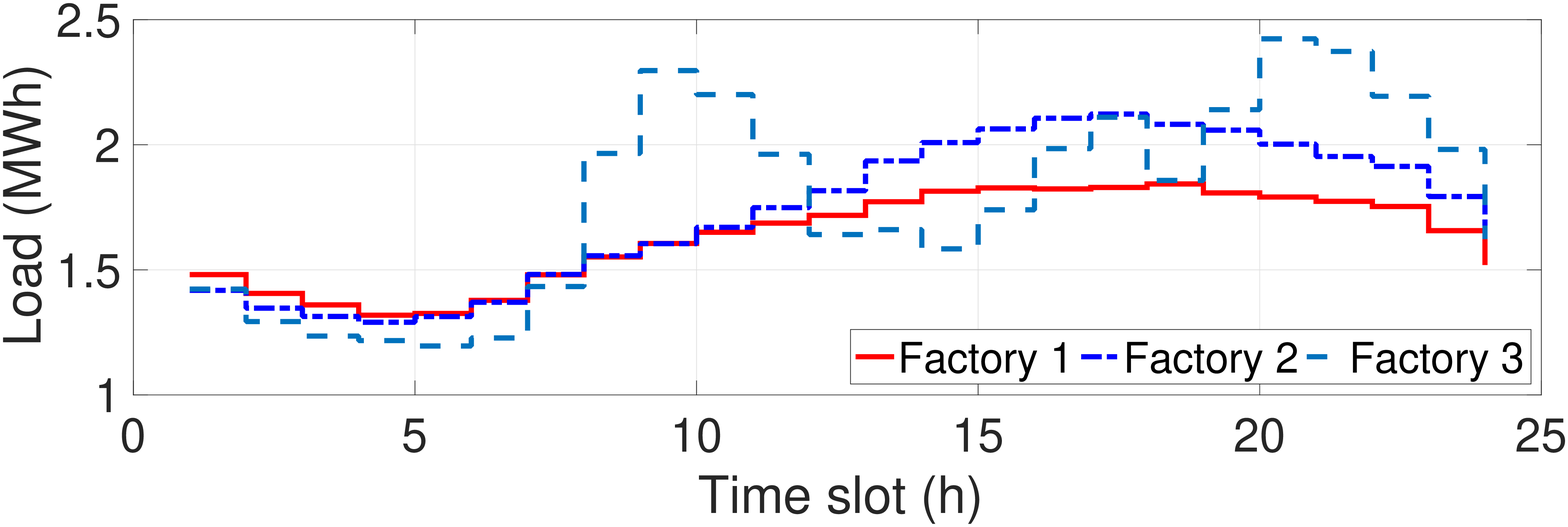}}
  \centerline{\scriptsize{{(b) Electrical load of factories }}}
\end{minipage}
\caption{Data from website.}
\label{fig2}
\end{figure}

Three different cases are used for comparison to evaluate the proposed algorithm: {Case 1 is a stochastic optimization algorithm with two-timescale (denoted as TA), which is similar to the algorithm of \cite{Jia2011Multi}, where there is no energy storage. Case 2 is an algorithm based on stochastic gradient without renewable energy resource (denoted as OA), which is similar to the algorithm of \cite{Deng2014Load}. To verify the effectiveness of the incentive mechanism, Case 3, a censored version of the proposed algorithm (denoted as CA) without incentive mechanism, is utilized for comparison.}
Fig.~\ref{fig2} (a) shows the electricity price of JiangSu Electric Power Company \cite{Illinois}.  %Considering that the electricity price is dynamic, the electricity demand of factories will be mainly analyzed. %The simulation results of the three approaches in section 3 are shown in Figs.~\ref{fig2}(b)-(d). 
%The data of photovoltaic systems is provided by Renewables.ninja \cite{Institute}, as shown in Fig.~\ref{fig2} (b).  in a general industrial environment 
Fig.~\ref{fig2} (b) shows the hourly load of factories from PJM \cite{Wang2021Multi}. %The total time length of the data is 24 hours. 

%\begin{figure}
%\centering
%\begin{minipage}{\linewidth}
%\begin{minipage}{0.49\linewidth}
  %  \centerline{\includegraphics[width=0.49\hsize]{price.eps}}
%  \centerline{\scriptsize{(a) }}
%\end{minipage}
%\hspace{-5pt}
%\begin{minipage}{1\linewidth}
 % \centerline{\includegraphics[width=\hsize]{pv2.eps}}
%  \centerline{\scriptsize{(b) }}
%\end{minipage}
%\begin{minipage}{\linewidth}
%  \centerline{\includegraphics[width=0.49\hsize]{ed.eps}}
%  \centerline{\includegraphics[width=\hsize]{f123o1.eps}}
%  \centerline{\scriptsize{(b) }}
%\end{minipage}
%\begin{minipage}{1\linewidth}
%  \centerline{\includegraphics[width=\hsize]{toloc1.eps}}
%  \centerline{\includegraphics[width=\hsize]{f123c1.eps}}
%  \centerline{\scriptsize{(d) }}
%\end{minipage}
%\caption{Data of the industrial park. (a) Electricity Price. (b) Electricity demand. }%. (b) Renewable Energy. Available capability, total load and the load of factories in (b) Approach 1. (c) Approach 2. (d)Approach 3.}
%\label{fig2}
%\end{figure}

\subsection{Performance Verification}

%the total electricity demand of each factory is 26.9 MWh for a whole day. Factory 1 works during 8:00-24:00, factories 2 and 3 work during 8:00-21:00. At any time slot, these factories need at least 0.6 MW power to maintain the necessary energy supply, such as lighting and keeping a suitable temperature.

%The ILs and ELs of each user follow a Poisson process, whose expectation is $50$. 
In order to confirm the performance analyzed in Section 4, the following two situations need to be considered. Situation 1: When renewable energy is sufficient (assumed to be six times the original renewable energy), the industrial park will have excess energy to sell. In this situation, the adjusted sales price ${p_{o}(t)}$ is proportional to the purchase price ${p_{e}(t)}$. Table 2 shows that the smaller the $\frac{p_{e}(t)}{p_{o}(t)}$, the smaller the total cost. %This is consistent with Lemma 2 and Theorem 1, where the cost gap increases with $\rho$. %The situation happens when the renewable energy sources are sufficient at part of a day, otherwise the industrial park has not extra energy to sell, the variation of $\frac{p_{e}}{p_{o}}$ is nonsense owing to the determined purchase price.
Situation 2: In current industrial production, there is no surplus energy in the industrial park to sell to the public utility company, and renewable energy cannot meet the total demand of factories at all time. In this situation, the purchase price ${p_{e}(t)}$ is adjusted. Table 2 shows that the smaller $p_{e,max}-p_{o,min}$, the smaller the total cost. % 
{These two situations are consistent with Lemma 2 and Theorem 1, where the cost gap increases with $\rho$.}

\begin{table}\footnotesize
\setlength\tabcolsep{2pt}
\centering
\caption{{The impact of the price variance on the total cost (\textyen)}}
%\begin{tabular}{m{1.4cm}|m{6.2cm}}
\begin{tabular}{|l| l l l l l l l l l l l |}
\hline
$\frac{p_{e}(t)}{p_{o}(t)}$ & 2 & 1.9 & 1.8 & 1.7 & 1.6 & 1.5 & 1.4 & 1.3 & 1.2 & 1.1 & 1 \\
\hline
Cost ($* 10^3$) & 94.1 & 93.4 & 92.7 & 92.0 & 91.2 & 89.6 & 88.2 & 86.7 & 84.7 & 81.9 & 78.4   \\%这一行改完了
%Cost  & 92.7983 & 92.2583 & 91.6253 & 90.4695 & 89.7674 & 89.0542 & 87.5163 & 85.5870 & 83.2263 & 80.2458 & 77.1188   \\
\hline
\end{tabular}
 \\ [5pt]
\begin{tabular}{|l| l l l l l l l l l |}
\hline
$p_{e,max}-p_{o,min}$ & 0.9 & 0.8 & 0.7 & 0.6 & 0.5 & 0.4 & 0.3 & 0.2 & 0.1 \\
\hline
Cost ($* 10^3$) & 141.5 & 135.9 & 130.0 & 123.5 & 116.4 & 107.3 & 97.5 & 86.7 & 75.2   \\%这一行改完了
%137.9291,132.0079,126.4438,120.5399,112.9075,104.4288,95.1254,85.0661,73.8904
\hline
\end{tabular}
\label{TAB1}
\end{table}

%\begin{figure}[htb]
%  \centerline{\includegraphics[width=\hsize]{y1.eps}}
%\caption{The impact of the price ratio $\frac{p_{e}(t)}{p_{o}(t)}$ on the total cost of the industrial park under sufficient renewable energy resources.}
%\label{fig8}
%\end{figure}

%\begin{figure}[htb]
%  \centerline{\includegraphics[width=\hsize]{y41.eps}}
%\caption{The impact of the price difference $p_{e,max}-p_{o,min}$ on the total cost of the industrial park under insufficient renewable energy resources.}
%\label{fig9}
%\end{figure}

\begin{figure}
\centering
\begin{minipage}{1\linewidth}
  \centerline{\includegraphics[width=.7\hsize]{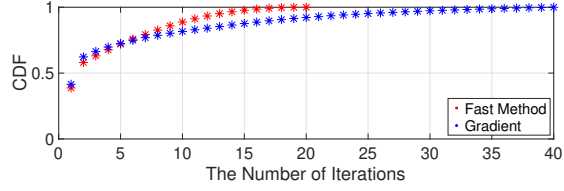}}
  \centerline{\scriptsize{(a) The comparison of CDF about the number of iterations. }}
\end{minipage}
\begin{minipage}{1\linewidth}
  \centerline{\includegraphics[width=.7\hsize]{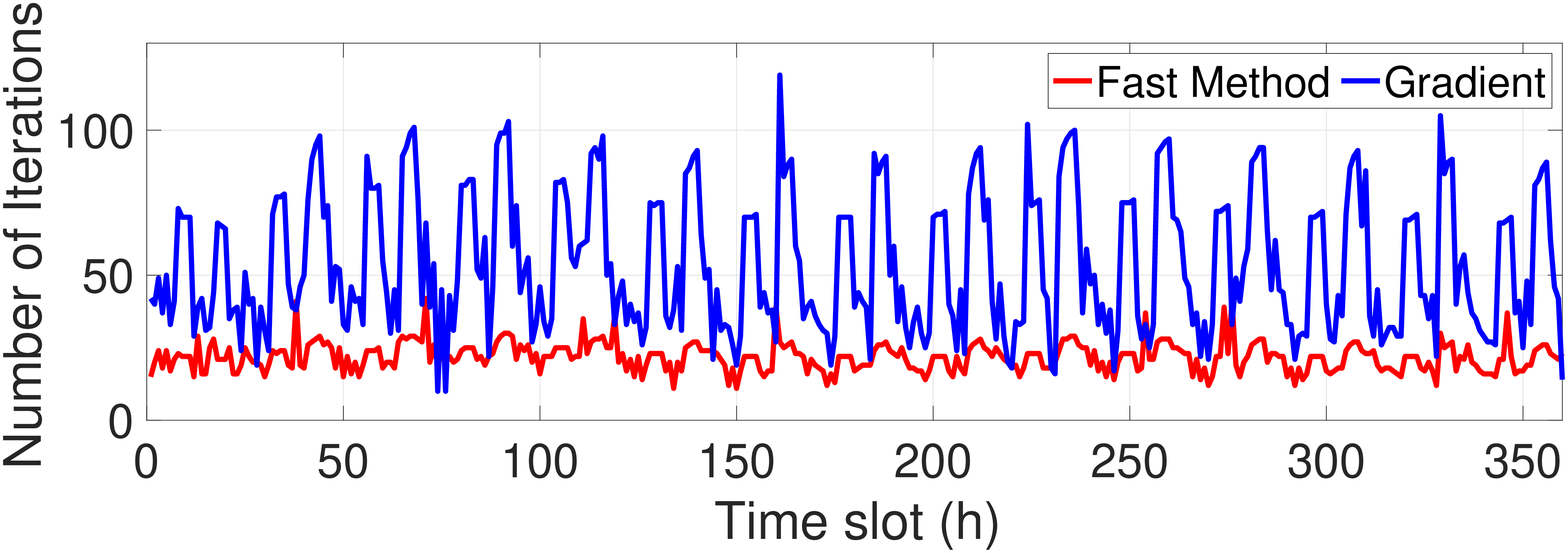}}
  \centerline{\scriptsize{(b) The number of iterations for different methods across 360 time slots. }}
\end{minipage}
\caption{{Comparison for different methods.} }
\label{fig3}
\end{figure}

The following simulation settings follow the previous subsection. %
The convergence of the solution to problem (\ref{micro17}) is shown in {Table 3 and Fig.~\ref{fig3}.  Table 3 shows the number of iterations for different methods during 12 hours. } The comparison of the cumulative distribution function (CDF) about the number of iterations for different methods at $t=1$ is shown in Fig.~\ref{fig3} (a). Then, the comparison time is increased to $T=360$ slots to further illustrate the results, and each time slot can only consider up to 200 mini-slots. The iteration will not be stopped until the difference between two iterations is less than 0.01 or the number of iteration for (\ref{micro17}) is greater than 200. Both the fast method (\ref{micro27}) and the conventional gradient algorithm (\ref{micro21}) have an iterative stepsize of $\sigma=0.2$. According to Fig.~\ref{fig3} (b), the iteration number of the fast method is about 20, while the conventional gradient algorithm requires more than 40 iterations to converge in most cases. %
{Therefore, by combining two previous iterations, the fast method reduces the undesired fluctuations of the gradient iteration to realize rapid convergence. }

\begin{table}\footnotesize
\setlength\tabcolsep{2pt}
\centering
\caption{{The number of iterations for different methods during 12 hours}}
%\begin{tabular}{m{1.4cm}|m{6.2cm}}
\begin{tabular}{|l |l l l l l l l l l l l l|}
\hline
$t (h)$ & 1 & 2 & 3 & 4 & 5 & 6 & 7 & 8 & 9 & 10 & 11 & 12\\
\hline
Fast Method& 15&20&24&18&24&17&21&23&22&22&22&15   \\
\hline
Gradient& 42&40&49&37	&50&33&41&73&70&70&70&29   \\
%Cost  & 92.7983 & 92.2583 & 91.6253 & 90.4695 & 89.7674 & 89.0542 & 87.5163 & 85.5870 & 83.2263 & 80.2458 & 77.1188   \\
\hline
\end{tabular}
\label{TAB3}
\end{table}

%Fig.~\ref{fig3} shows that the iterative reduction for the fast iterative algorithm is more than that for the standard dual gradient algorithm. 

%Fig.~\ref{1} shows the comparison between the optimal solution of the proposed algorithm and real optimal solution. The simulation results indicate that the gap between the performance of the proposed algorithm and optimal performance decreases as the stepsize $\rho$ decreases at the price of slow convergence, which is in accordance with the characterization of performance analyzed in Theorem 1.

\begin{figure}
\centering
%\begin{minipage}{1\linewidth}
%  \centerline{\includegraphics[width=\hsize]{tc2.eps}}
 % \centerline{\scriptsize{(a) }}
%\end{minipage}
\begin{minipage}{1\linewidth}
  \centerline{\includegraphics[width=\hsize]{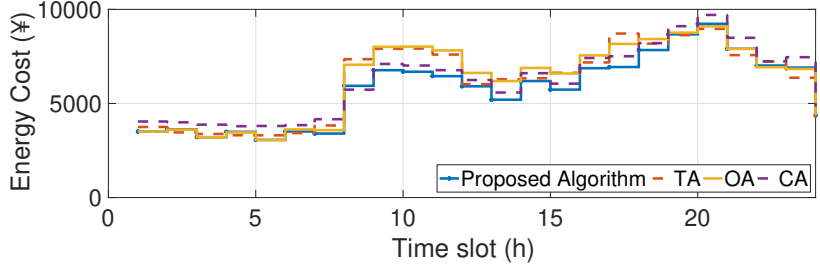}}
%  \centerline{\scriptsize{(b) }}
\end{minipage}
\caption{{Comparison of costs across 24 time slots for different methods.} }%(a) Total costs. (b) Costs across 24 time slots.}
\label{fig4}
\end{figure}

{Table 4 shows the comparison of the total costs for different methods. Fig.~\ref{fig4} shows the comparison of the costs across 24 time slots for different methods. The cost of the proposed algorithm is lower than that of TA, OA and CA. Due to the lack of energy storage, TA needs to purchase more electricity instead of  discharging during peak periods and is sensitive to the high electricity price and renewable energy consumption. Due to the lack of incentive mechanism and EL, CA is sensitive to changes in prices, renewable energy and loads. Therefore, the park needs to pay higher costs under TA/CA. The effect of OA approaches the proposed algorithm during 0:00-6:00 and 20:00-24:00, but it needs to buy more energy in daytime due to the lack of renewable energy in OA. Since TA, OA and CA need to purchase more high-priced electricity from the grid during peak demand, their costs are higher. The comparison of total costs between the proposed algorithm and TA shows that the cost is improved 6.17\% by the energy storage, and the comparison of total costs between the proposed algorithm and OA shows that the cost is improved 7.23\% through introducing renewable energy resource, and the comparison of total costs between the proposed algorithm and CA shows that the cost is improved 8.16\% by incentive mechanism. } Therefore, the proposed fast distributed algorithm takes full advantage of incentive mechanism and the characteristic of loads to coordinate multi-energy, and utilizes batteries, water tanks and distributed energy generation to minimize the total cost of the industrial park online.

\begin{table}\footnotesize
\centering
\caption{{Comparison of total costs for different methods (\textyen)}}
%\begin{tabular}{m{1.4cm}|m{6.2cm}}
\begin{tabular}{|l |l l l l|}
\hline
Methods & Proposed algorithm & TA & OA & CA  \\
\hline
Cost ($* 10^3$) & 138.5 & 147.6 & 149.3 & 150.8  \\
\hline
\end{tabular}
\end{table}

\begin{figure}
\centering
\begin{minipage}{1\linewidth}
  \centerline{\includegraphics[width=0.7\textwidth]{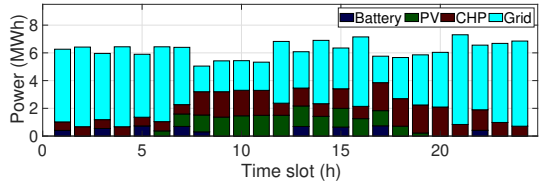}}
  \centerline{\scriptsize{(a) Power generation/discharging}}
\end{minipage}
%\hspace{-5pt}
\begin{minipage}{1\linewidth}
 \centerline{\includegraphics[width=0.7\textwidth]{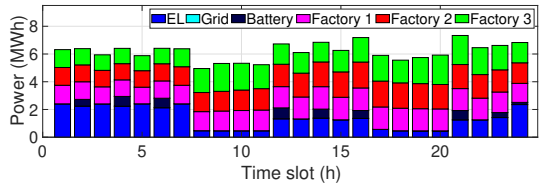}}
  \centerline{\scriptsize{(b) Power consumption/charging }}
\end{minipage}
\caption{{Power distribution profiles of the park. }}
\label{fig5}
\end{figure}

\begin{figure}
\centering
\begin{minipage}{1\linewidth}
  \centerline{\includegraphics[width=0.7\textwidth]{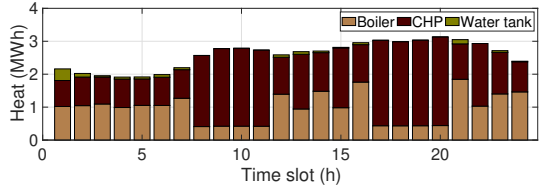}}
  \centerline{\scriptsize{(a) Heat generation/discharging}}
\end{minipage}
%\hspace{-5pt}
\begin{minipage}{1\linewidth}
 \centerline{\includegraphics[width=0.7\textwidth]{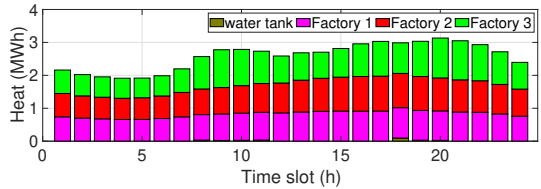}}
  \centerline{\scriptsize{(b) Heat consumption/charging }}
\end{minipage}
\caption{{Heat distribution profiles of the park. }}
\label{fig6}
\end{figure}

\begin{figure}
\centering
\begin{minipage}{1\linewidth}
  \centerline{\includegraphics[width=0.7\textwidth]{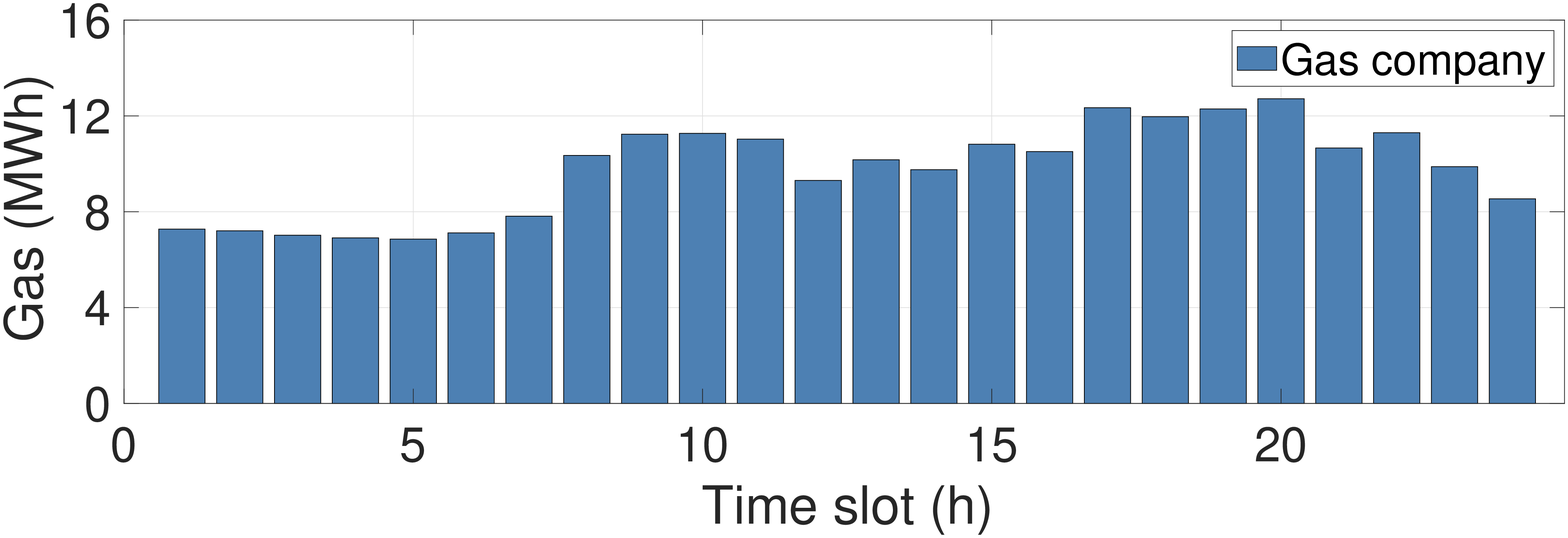}}
  \centerline{\scriptsize{(a) Gas purchase}}
\end{minipage}
%\hspace{-5pt}
\begin{minipage}{1\linewidth}
 \centerline{\includegraphics[width=0.7\textwidth]{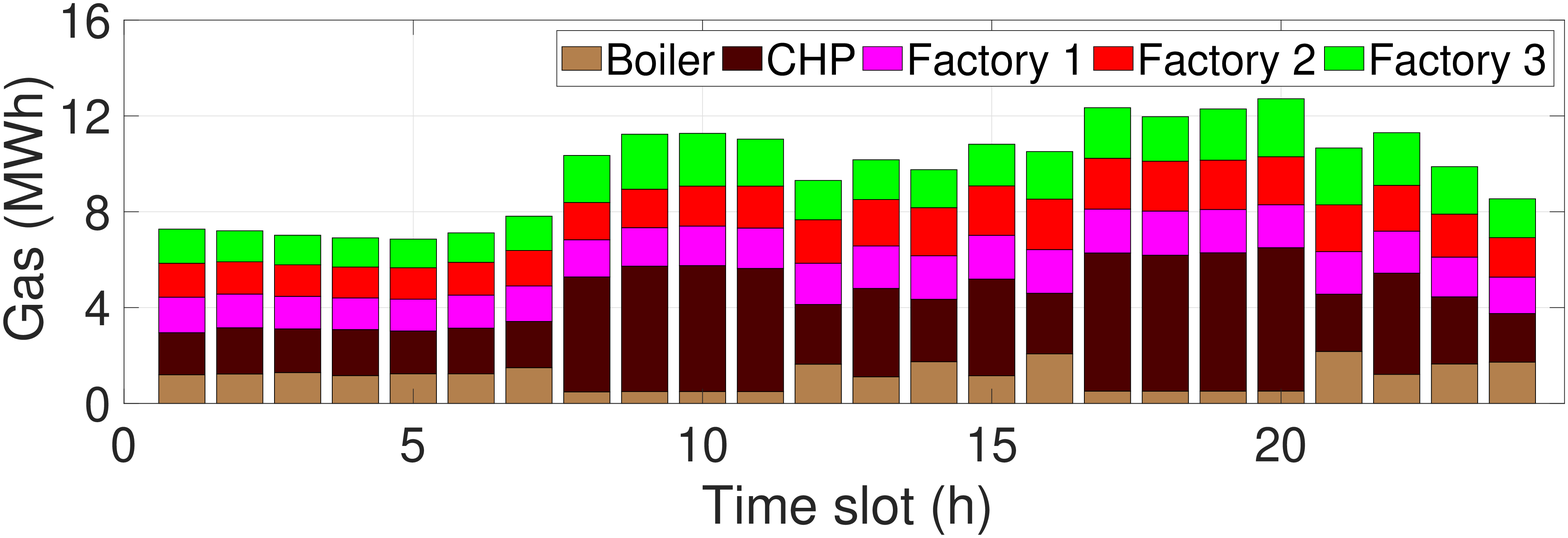}}
  \centerline{\scriptsize{(b) Gas consumption }}
\end{minipage}
\caption{{Gas distribution profiles of the park. }}
\label{fig7}
\end{figure}

Figs. ~\ref{fig5}-\ref{fig7} show the energy generation/discharging and consumption/charging of the park, where the data of photovoltaic systems is provided by Renewables.ninja \cite{Institute}. As the price of electricity sold by the power company changes over time, the CHP units are introduced to output electricity rather than buying the high-priced electricity, and supplying heat to the industrial park at the same time, as shown in Figs. ~\ref{fig5}-\ref{fig7}. Fig.~\ref{fig5} shows that the batteries are discharged when the electricity price is high, such as 8:00, 9:00 and 17:00, and charged when the electricity price is low, such as 12:00, 14:00 and 21:00. In addition, according to Fig.~\ref{fig5}, users have less EL under high electricity price. When the heat generated by CHP units cannot meet the demand, the boilers meet the unmet demand, as shown in Figs.~\ref{fig6} and \ref{fig7}. Figs.~\ref{fig5}-\ref{fig7} show that the proposed fast distributed algorithm achieves energy transactions with power and gas companies, multi-energy demand shift, and flexible supply of energy through energy storage. Therefore, the proposed algorithm can realize the time-average cost that is close to the optimal cost and ensure real-time coordination and fast convergence.

%\begin{figure}
%\centering
%\begin{minipage}{1\linewidth}
%  \centerline{\includegraphics[width=.49\hsize]{eg311.eps}}
%  \centerline{\scriptsize{(a) }}
%\end{minipage}
%\begin{minipage}{1\linewidth}
%  \centerline{\includegraphics[width=.49\hsize]{ec311.eps}}
%  \centerline{\scriptsize{(b) }}
%\end{minipage}
%\caption{Electricity profiles of the industrial park using the proposed algorithm. (a) Electricity generation. (b) Electricity consumption.}
%\label{fig5}
%\end{figure}

%\begin{figure}
%\centering
%\begin{minipage}{1\linewidth}
%  \centerline{\includegraphics[width=\hsize]{hg3.eps}}
%  \centerline{\scriptsize{(a) }}
%\end{minipage}
%\begin{minipage}{1\linewidth}
%  \centerline{\includegraphics[width=\hsize]{hc3.eps}}
%  \centerline{\scriptsize{(b) }}
%\end{minipage}
%\caption{Heat profiles of the industrial park using the proposed algorithm. (a) Heat generation. (b) Heat consumption.}
%\label{fig6}
%\end{figure}

%\begin{figure}
%\centering
%\begin{minipage}{1\linewidth}
%  \centerline{\includegraphics[width=\hsize]{gg3.eps}}
%  \centerline{\scriptsize{(a) }}
%\end{minipage}
%\begin{minipage}{1\linewidth}
%  \centerline{\includegraphics[width=\hsize]{gc3.eps}}
%  \centerline{\scriptsize{(b) }}
%\end{minipage}
%\caption{Gas profiles of the industrial park the proposed algorithm. (a) Gas generation. (b) Gas consumption.}
%\label{fig7}
%\end{figure}

\section{Conclusion}
In this paper, the energy management problem of a multi-energy industrial park is investigated, which is an imperative issue of today's industrial production. A systematic online energy cost minimization framework composed of energy hubs and users is presented, which makes full use of the adjustment mechanism of elastic load, the compensation mechanism of inelastic load and the coordination mechanism of multi source-load-storage. An incentive mechanism is implemented by estimating users' willingness to shift peak loads without knowing information of each individual user. The energy scheduling issue is constructed as a two-timescale optimization problem, which can further consider the temporal and spatial changes of renewable energy, load and electricity prices, to achieve energy storage balancing and real-time load balancing while respecting energy constraints. A fast distributed algorithm based on two-stage dual decomposition is proposed to deal with temporally-coupled constraints and ensure real-time coordination of instantaneous scheduling.
Finally, the performance and feasibility of the proposed algorithm are verified by theoretical analysis and case studies. 

In this paper, an industrial park consists of two energy hubs and three factories. According to actual industrial scenarios, a park can have many generation plants and factories. Therefore, the scale and characteristics of the park need to be further considered. Investigating some diversified multi-energy management frameworks, such as Ref. \cite{Zhou2021Distributionally}, to solve the practical problems of multi-energy market is an important line of inquiry. Another line is to design some management schemes, like Ref. \cite{Li2021Optimal}, to further explore the deployment of battery and thermal energy storage to improve energy efficiency.

\begin{appendix}
%\section{}
%\begin{appendices}
\section{Proof of Lemma 2}

{ The induction is used to prove Lemma 2. First, the conditions hold at time 0 and still hold at time $t$. Then, the following three cases are considered.}
\begin{enumerate}
\item Case 1: $\lambda_{ke}(t)\in [-p_{o,min}, \rho B_{k,max}-\rho B_{k,min}-p_{e,max}-\rho D_{ke,max}]$, where $p_{o,min}=\min\{p_{o}(t), \forall t\}$, and $p_{e,max}=\max\{p_{e}(t), \forall t\}$. In this case, according to Lemma 2, $C_{ke}(t)=0$ and $D_{ke}(t)=D_{ke,max}$. Since $-p_{e,max}<-p_{o,min}$, $\lambda_{ke}(t+1)=\lambda_{ke}(t)+\rho (C_{ke}(t)-D_{ke}(t))\in [-p_{e,max}-\rho D_{ke,max}, \\\rho B_{k,max}-\rho B_{k,min}-p_{e,max}-\rho D_{ke,max}]$.

\item Case 2: $\lambda_{ke}(t)\in [-p_{e,max}, -p_{o,min}]$. In this case, since $\rho\geq\rho_{min}$, $\lambda_{ke}(t+1)=\lambda_{ke}(t)+\rho (C_{ke}(t)-D_{ke}(t))\in [-p_{e,max}-\rho D_{ke,max},-p_{o,min}+\rho C_{ke,max}] \subseteq [-p_{e,max}-\rho D_{ke,max}, \rho B_{k,max}-\rho B_{k,min}-p_{e,max}-\rho D_{ke,max}]$.

\item Case 3: $\lambda_{ke}(t)\in [-p_{e,max}-\rho D_{ke,max}, -p_{e,max}]$. In this case, according to Lemma 2, $C_{ke}(t)=C_{ke,max}$ and $D_{ke}(t)=0$. Since $-p_{e,max}<-p_{o,min}$ and $C_{ke,max}>0$, $\lambda_{ke}(t+1)=\lambda_{ke}(t)+\rho (C_{ke}(t)-D_{ke}(t))\in [-p_{e,max}-\rho D_{ke,max}+\rho C_{ke,max}, -p_{e,max}+\rho C_{ke,max}]\subseteq [-p_{e,max}-\rho D_{ke,max}+\rho C_{ke,max}, \\-p_{o,min}+\rho C_{ke,max}]\subseteq [-p_{e,max}-\rho D_{ke,max}, \rho B_{k,max}-\rho B_{k,min}-p_{e,max}-\rho D_{ke,max}]$.
\end{enumerate}

It is similar to analyze $\lambda_{kh}(t)$.
\end{appendix}

%\section{Bibliography styles}

%There are various bibliography styles available. You can select the style of your choice in the preamble of this document. These styles are Elsevier styles based on standard styles like Harvard and Vancouver. Please use Bib\TeX\ to generate your bibliography and include DOIs whenever available.

%Here are two sample references: \cite{Feynman1963118,Dirac1953888}.

%\section*{References}

\bibliography{mybibfile}

\end{document}